\DeclareUnicodeCharacter{2500}{}
\documentclass[10pt]{article}
\usepackage{xurl} % for URLs in bibliography
\usepackage{cite} % for grouped numeric citations (optional)

\usepackage[utf8]{inputenc}
\usepackage[T1]{fontenc}
\usepackage{geometry}
\usepackage{lmodern,microtype}
\usepackage{amsmath,amssymb,amsfonts,amsthm,mathtools}
\usepackage{bm,bbm}
\usepackage{graphicx}
\usepackage{booktabs}
\usepackage{enumitem}
\usepackage{algorithm}
\usepackage{algpseudocode}
\usepackage{hyperref}
\usepackage[nameinlink,capitalize]{cleveref}
\usepackage{tikz-cd}
\usepackage{subcaption} % in preamble
\usepackage[font=small,labelfont=bf]{caption}
\usepackage[sort&compress,numbers]{natbib}
\usepackage{xcolor}   % For comments
\usepackage{lineno}   % Line numbers for review
\usepackage{amsmath, amssymb, amsthm, mathrsfs}

\usepackage{amsmath}
\usepackage{amsfonts}
\usepackage{amssymb}
\usepackage{geometry}
\usepackage{cite}
\usepackage{graphicx} % Added for potential figure handling if needed
\usepackage{url}      % For handling emails/links
\usepackage[utf8]{inputenc}
\usepackage[T1]{fontenc}
\usepackage{amsmath, amssymb, amsfonts} 
\usepackage{booktabs}                   % For professional table lines (\toprule, \midrule, \bottomrule)
\usepackage{multirow}                   % For merging cells in the Dataset table
\usepackage[table]{xcolor}              % For row colors (\rowcolor)
\usepackage{graphicx}                   % For including your .png figures
\usepackage{caption}                    % For better control over figure/table captions
\usepackage{cite}                       % For managing your bibliographic citations
\usepackage{hyperref}                   % For clickable links and cross-references
\usepackage{longtable}
\usepackage{booktabs}
\usepackage{amsmath}
\usepackage[utf8]{inputenc}
\usepackage{booktabs}   % For professional looking tables
\usepackage{multirow}   % For merging rows
\usepackage[table]{xcolor} % For coloring rows
\usepackage{array}      % For better column definitions

\definecolor{headerbeige}{RGB}{220, 222, 210}
\definecolor{finetunegreen}{HTML}{B0D7AA} %

\usepackage{amsmath}
\usepackage{amssymb}
\usepackage{algorithm}
\usepackage{algpseudocode}
\usepackage{bm}
\usepackage{booktabs}
\usepackage{caption}
\title{TopoLIA: Topology-Inspired Latent Interaction Attention  for Drug Discovery}
 
 \author{Xiang Liu$^{1,\dagger}$,  Yiming Ren$^{1,\dagger}$, Mustafa Hajij$^2$, Pietro Li\`{o}$^{3}$, and Guo-Wei Wei$^{1,4,5,*}$
 \\[2mm]
        \\    % Author name
 		\\
         $^1$Department of Mathematics,\\
 		  University of Georgia, Athens, GA 30602, USA. \\
         $^2$Department of Data Science, University of San Francisco, CA 94117, USA.\\
         $^3$ Department of Computer Science and Technology, \\ 
         Cambridge University, Cambridge, United Kingdom. \\
        $^4$Department of Biochemistry and Molecular Biology,\\
 		University of Georgia, Athens, GA 30602, USA.   \\
         $^5$School of Computing,\\
 		University of Georgia, Athens, GA 30602, USA.   
 	}
\date{}
\begin{document}

\maketitle

\begingroup
\renewcommand{\thefootnote}{\fnsymbol{footnote}}
\footnotetext[2]{The first two authors contributed equally to this work.}
\footnotetext[1]{Corresponding author.   Email: guowei.wei@uga.edu}
\endgroup

\begin{abstract}

The growing sophistication of drug discovery calls for new advances in artificial intelligence (AI). Transformer architecture has underpinned many of the recent breakthroughs in AI, yet its conventional source-target attention mechanism can become a bottleneck for modeling complex interactions. Here, we introduce TopoLIA, a topology-inspired latent interaction attention mechanism that learns not only how much information should be exchanged between interaction partners, but also how that information should be transformed. Source-target-dependent attention is mediated through a compact latent interaction space, providing a structured representation of complex interactions. For complex data, we further incorporate topological spectral representations to capture multiscale data geometry and extract algebraic relationships. TopoLIA thus incorporates topological principles into both data representation and neural network architecture. We demonstrate TopoLIA on protein--ligand binding, where persistent spectra encode multiscale topological and geometric patterns in three-dimensional (3D) protein--ligand complexes, while latent interaction attention (LIA) enables structured, source-target-dependent information exchange. We pretrain TopoLIA by reconstructing masked multiscale spectral tokens from unlabeled molecular structures, yielding transferable structural representations without task-specific annotations. Across multiple scoring and docking benchmarks, TopoLIA achieves state-of-the-art performance and consistently outperforms its standard-attention counterpart. Systematic rank analyses further show that compact latent dimensions yield the best performance, supporting compact interaction spaces as an effective inductive bias for biomolecular learning. More broadly, LIA provides a general framework for structured, source-target-dependent information exchange that is applicable to diverse scientific and AI problems involving complex heterogeneous interactions.

\end{abstract}

\newpage

\section{Introduction}
Accelerating drug discovery remains a major challenge in modern healthcare, yet the transition from drug candidate identification to clinical approval often requires more than a decade and billions of dollars \cite{scannell2012diagnosing}. The limitations of traditional methodologies, including molecular docking, free energy perturbation, and empirical modeling, have motivated the development of deep learning approaches \cite{lo2018machine, ain2015machine, lin2023evolutionary}. However, conventional deep learning architectures may struggle to represent the non-Euclidean, irregular, and multiscale geometry of 3D protein--ligand complexes. Although transformer models have established a powerful paradigm for representation learning \cite{vaswani2017attention, devlin2019bert, ouyang2022training}, standard architectures do not explicitly encode the high-order geometric relationships present in biomolecular systems. More fundamentally, standard attention primarily captures source--target dependencies through scalar attention weights, while the transformed value of each source remains independent of the specific target.

To better capture the structural complexity of biomolecular systems, advanced mathematical frameworks have increasingly been explored in biomolecular machine learning \cite{cang2018representability, meng2021persistent, nguyen2019agl, feng2025caml, liu2021persistent}. Persistent homology (PH) \cite{edelsbrunner2002topological, zomorodian2004computing} has successfully reduced complex biological data to robust topological signatures \cite{cang2018integration}. Persistent spectral theory (PST) \cite{wang2020persistent} extends PH beyond topological invariants by incorporating non-harmonic spectra that capture subtle geometric and shape-dependent variations across multiple scales \cite{wei2025persistent}. These topological representations further motivated the development of topological deep learning (TDL) in 2017 \cite{cang2017topologynet}, establishing topology-aware learning as an increasingly important direction in molecular science \cite{wee2025review}. However, most existing TDL models use topology primarily for feature extraction rather than incorporating it directly into neural architectures.

More recently, topological principles have begun to directly shape neural network architectures \cite{papamarkou2024position}. Copresheaf topological neural networks (CTNNs) \cite{hajij2022topological,hajij2025copresheaf} provide one such example. Rather than assuming a single global latent space, CTNNs associate local regions with their own feature spaces and mediate information exchange between them through learnable copresheaf structure maps. This perspective suggests that information exchange should involve not only how much information is passed, but also how that information is transformed across different representations. However, directly implementing such interaction-dependent transformations in a high-dimensional feature space can be computationally expensive and may introduce unnecessary flexibility. We therefore hypothesize that biomolecular information exchange can be effectively mediated through a much lower-dimensional latent interaction space than the original feature space.

Inspired by this perspective, we introduce latent interaction attention (LIA), a structured attention mechanism in which information exchange is mediated through a learned low-dimensional interaction space. In standard attention, pairwise relationships primarily determine how much information is exchanged from each source through scalar attention coefficients, while the same source representation is shared across different targets. LIA additionally learns how the source value should be transformed for different targets during information exchange. Specifically, each source representation is first projected into a compact latent interaction space through a source-dependent transformation, and the resulting latent representation is then mapped to the target space through a target-dependent transformation. Their composition thus defines a structured source-to-target transformation through the latent interaction space, enabling target-dependent information transformation without explicitly parameterizing a separate high-dimensional mapping for every source--target pair. This low-dimensional bottleneck provides a structured inductive bias for heterogeneous biomolecular interactions while controlling the computational cost of interaction-dependent transformations.

Building on LIA, we develop TopoLIA by integrating latent interaction attention with persistent spectral representations and self-supervised pretraining for learning three-dimensional protein--ligand interactions. Persistent spectra encode the multiscale topological and geometric organization of protein--ligand complexes, while LIA enables these heterogeneous structural representations to exchange information through compact latent interaction spaces using source- and target-dependent projections. To exploit large collections of molecular structures without requiring task-specific annotations, we further employ a self-supervised masked pretraining strategy \cite{devlin2019bert, ouyang2022training, he2022masked}. The model reconstructs masked multiscale spectral tokens from their molecular context, encouraging it to learn dependencies among atomic composition, spatial organization, topological cavities, and non-harmonic geometric variations. The resulting structural representations can subsequently be transferred to downstream molecular tasks. Across multiple protein--ligand benchmarks, TopoLIA achieves state-of-the-art performance in both scoring and docking tasks and consistently outperforms its standard-attention counterpart. Systematic rank analyses further support the low-dimensional interaction hypothesis, with compact latent dimensions consistently yielding the best performance.

\section{Results}

\subsection{Overview of TopoLIA}

\begin{figure}[htp] 
\centering  
\includegraphics[width=1.0\linewidth]{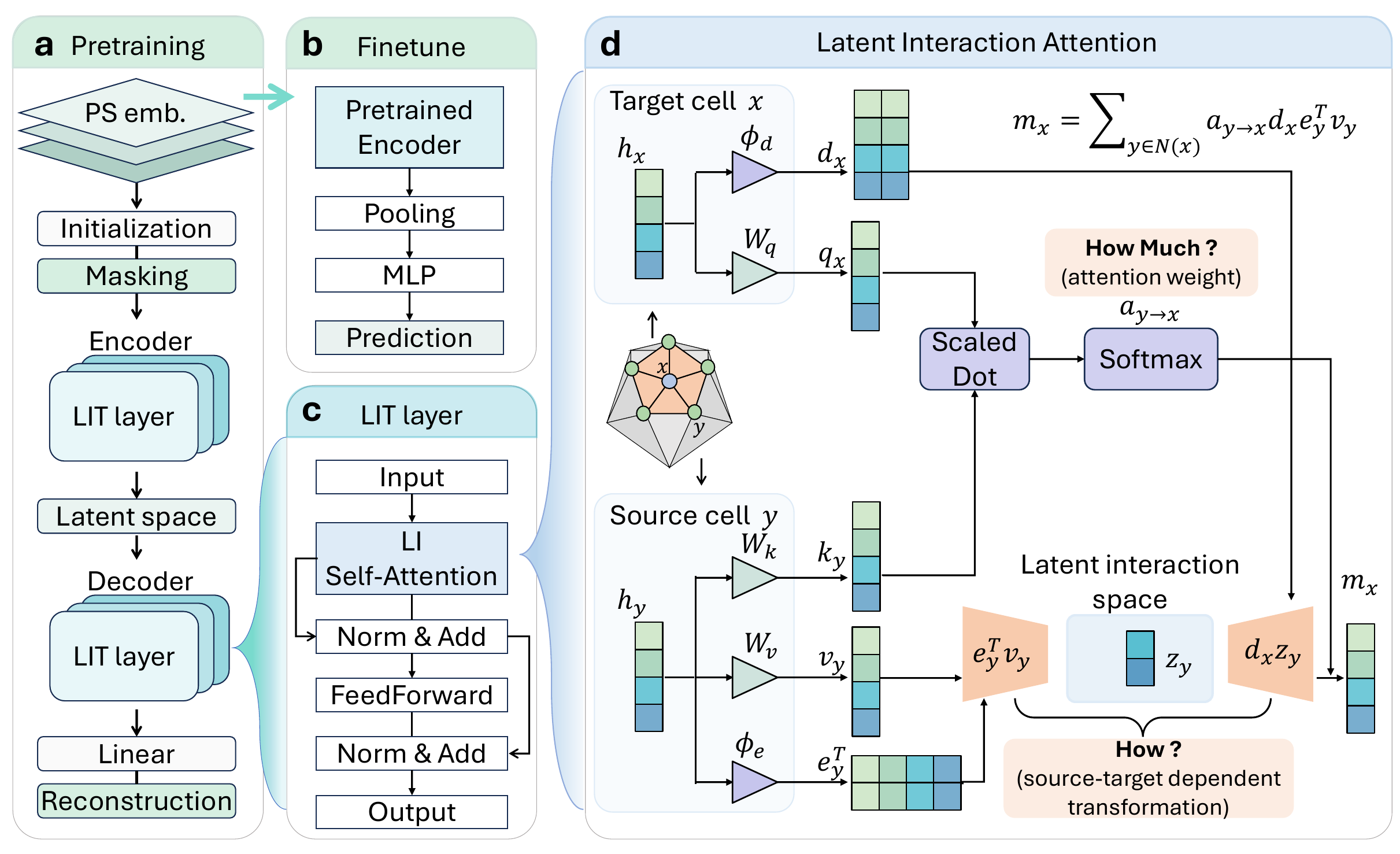} 
\caption{Overview of the TopoLIA framework. 
(a): Self-supervised pretraining: multiscale persistent spectral embeddings (PS emb.) are used as structural tokens for masked self-supervised pretraining. A subset of the tokens is masked and reconstructed using an encoder-decoder architecture composed of latent interaction transformer (LIT) layers.
(b): Downstream task fine-tuning: the pretrained encoder is transferred to downstream molecular tasks, where the learned representations are aggregated through pooling and mapped to task-specific predictions using a multilayer perceptron (MLP).
(c): Architecture of a LIT layer: each LIT layer preserves the overall transformer block structure while replacing standard self-attention with latent interaction attention (LIA).
(d): Latent interaction attention: for a target $x$ and source $y$, query-key similarity determines the scalar attention weight $a_{y\to x}$, specifying how much information is exchanged. In parallel, the source value $v_y$ is projected by the source-dependent map $e_y^T$ into a compact latent interaction space and subsequently transformed through the target-dependent projection $d_x$, specifying how the information is transformed. The resulting source messages are weighted and aggregated to obtain the target representation $m_x$.
}
\label{fig:workflow} 
\end{figure}

TopoLIA integrates multiscale persistent spectral representations, topology-inspired latent interaction attention, and self-supervised pretraining for protein--ligand interaction learning (Fig. \ref{fig:workflow}). At the representation level, persistent spectral embeddings encode multiscale topological and geometric information from three-dimensional protein--ligand complexes and serve as structural tokens for masked self-supervised pretraining. During pretraining, a subset of these tokens is masked and reconstructed through an encoder-decoder architecture, encouraging the model to capture dependencies across complementary structural representations (Fig. \ref{fig:workflow}a). The pretrained encoder is subsequently transferred to downstream molecular tasks through fine-tuning, with its representations aggregated by pooling and mapped to task-specific outputs through a multilayer perceptron (Fig. \ref{fig:workflow}b).

At the neural architecture level, both the encoder and decoder are built from latent interaction transformer (LIT) layers, which preserve the overall transformer block structure while replacing standard self-attention with latent interaction attention (LIA) (Figure \ref{fig:workflow}c). LIA separates two complementary aspects of information exchange: how much information should be exchanged and how that information should be transformed (Figure \ref{fig:workflow}d). The attention weight determines how much information is exchanged, whereas the source-target-dependent transformation determines how information from a source is adapted to a target. In standard attention, a source-specific value representation is typically shared across targets, with source-target dependence entering primarily through scalar attention weights. LIA additionally makes the transformation of the transported information dependent on both its source and target through a compact latent interaction space. For biomolecular interactions, this compact latent interaction space reflects the hypothesis that effective biomolecular information exchange can be mediated through a substantially lower-dimensional interaction space than the full feature space.

\subsection{Datasets}
TopoLIA was evaluated on multiple benchmarks from PDBbind database \cite{liu2015pdb}. For 
self-supervised pre-training, we curated 19,462 non-overlapping, unlabeled protein--ligand complexes from PDBbind v.2007, v.2013, v.2016, and v.2020. For scoring power, TopoLIA was evaluated on CASF-2007, CASF-2013, and CASF-2016 core sets, with the corresponding refined sets as training after excluding core-set samples \cite{li2014comparative, cheng2009comparative, su2018comparative}. Docking power was evaluated on the CASF-2013 of 195 ligands, with additional GOLD-generated poses used for training\cite{nguyen2019agl}. Generalization over sequence similarity was assessed on the Atom3D-curated LBA30 and LBA60 datasets \cite{townshend2020atom3d}. Ten-fold cross-validation was performed on the PDBbind v.2020 refined set. These datasets and evaluation protocols are summarized in Table S1, with additional details provided in Section S1 of the Supplementary Information.

%More details are provided in Section S1 of the Supplementary Information. 

\subsection{Scoring Power on CASF Benchmarks}
For scoring power evaluation, TopoLIA was quantified through the Pearson Correlation Coefficient (PCC) and Root Mean Squared Error (RMSE). To ensure the statistical robustness of the scoring results, we performed ten independent runs using different random seeds. The final metrics reported for each benchmark are based on the average predictions across these ten iterations. While binding affinities in PDBbind are provided in units of $pK_d$, predicted values were multiplied by a factor of $-1.36$ for conversion to kcal/mol.

\begin{figure}[hp] 
\centering  
\includegraphics[width=\linewidth]{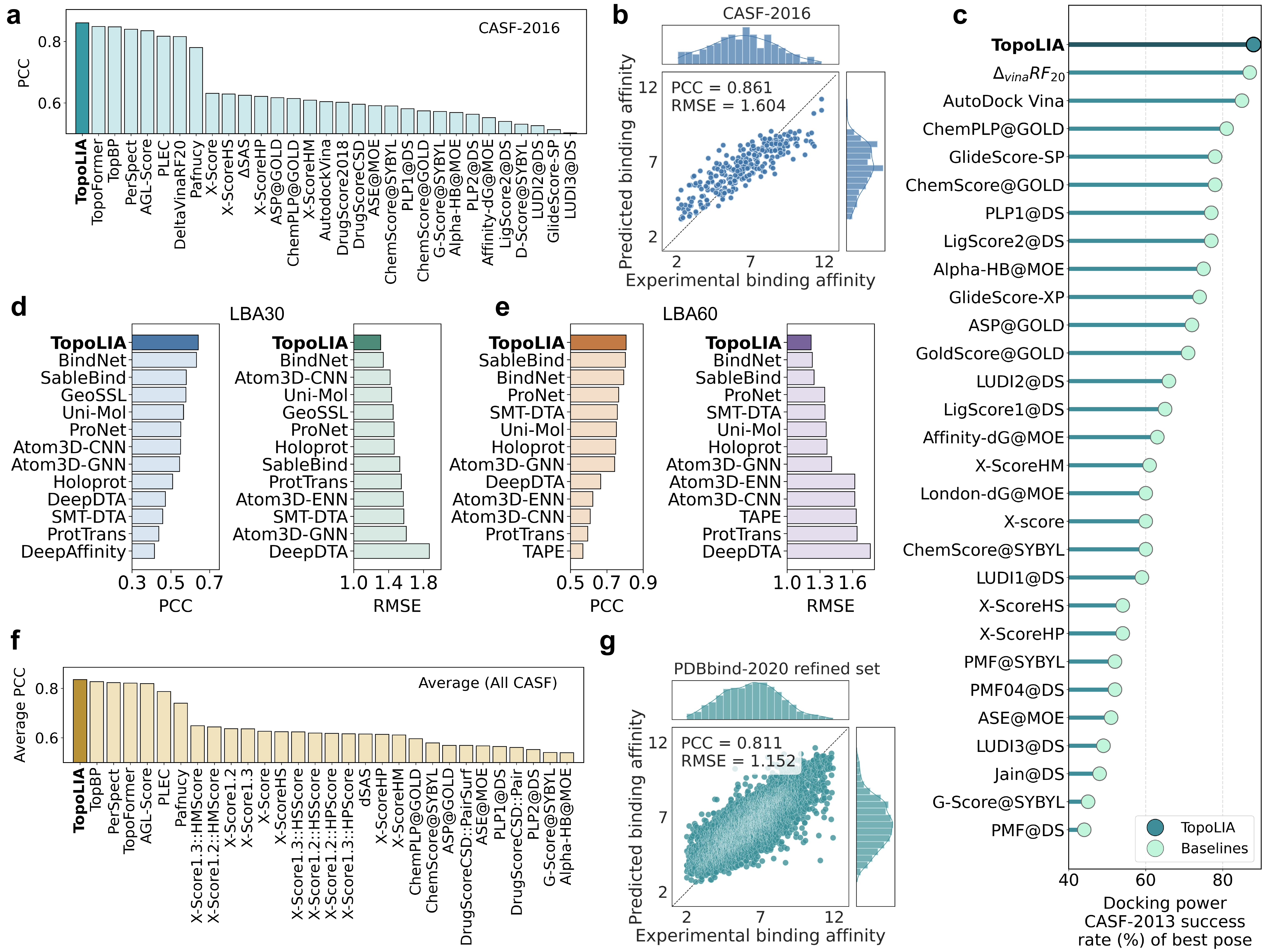} 
\caption{Comprehensive performance evaluation of the TopoLIA model for protein--ligand binding affinity scoring and docking tasks.
{(a)}: Comparative analysis of the PCC between the proposed TopoLIA model and established scoring functions \cite{stepniewska2018development, cang2018representability, meng2021persistent, li2014comparative, su2018comparative, cheng2009comparative, wang2017improving, nguyen2019agl, wojcikowski2019development} on the CASF-2016 core set. 
{(b)}: Scatter plot illustrating the correlation between experimental and predicted binding affinities for the CASF-2016 benchmark. 
{(c)}: Docking power evaluation on the CASF-2013 dataset, displaying the success rate of the top-ranked pose for TopoLIA against classical baselines \cite{cang2017topologynet,huang2006iterative,wang2017improving,cheng2009comparative,li2013id,li2015improving,nguyen2019dg}.
{(d, e)}: Generalization evaluation showcasing PCC and RMSE metrics on the sequence-identity split datasets, LBA30 and LBA60, respectively \cite{ozturk2018deepdta,rao2019evaluating,elnaggar2021prottrans,townshend2020atom3d,somnath2021multi,wang2022learning,karimi2019deepaffinity,pei2022smt,liu2022molecular,zhou2023uni,feng2023protein,li2025harnessing}. 
{(f)}: Average PCC achieved across top evaluated CASF benchmarks. 
{(g)}: Scatter plot detailing the robust correlation results from the rigorous 10-fold cross-validation on the structurally diverse PDBbind v.2020 refined set.} 
\label{fig:PCC} 
\end{figure}

TopoLIA demonstrates robust scoring performance across successive PDBbind datasets of increasing complexity. Overall, it achieves an industry-leading average performance across all evaluated datasets (Figure~\ref{fig:PCC}f), with a comprehensive quantitative comparison against representative state-of-the-art approaches provided in Table S4. On the CASF-2007 benchmark, TopoLIA was trained on a refined set of 1,105 complexes and evaluated on the standard 195-complex test set. It achieved a competitive PCC of 0.832, matching the leading methods presented in Figure S1a and showing a strong linear correlation between experimental data and predicted affinities in Figure S1b.

The scoring power was further validated using the CASF-2013 benchmark. Trained on a refined set of 2,764 complexes strictly excluding the 195 CASF-2013 core set entries, TopoLIA achieved a PCC of 0.814 and an RMSE of 1.908 kcal/mol. Notably, it outperformed previous persistent spectral approaches~\cite{meng2021persistent} and top-tier models like TopBP~\cite{cang2018representability}. This comparative advantage is visualized in Figure S1c, with the correlation between experimental and predicted values in Figure S1d.

The CASF-2016 benchmark represents the most extensive scoring assessment in this study, with 3772 training complexes and 285 test samples. This dataset has historically served as a rigorous testing ground for sophisticated mathematical and physical frameworks~\cite{ballester2010machine, wang2017improving, Li2023Development}. On this benchmark, TopoLIA achieved a PCC of 0.861, significantly surpassing leading architectures including Topoformer~\cite{chen2024multiscale} and PerSpect-ML ~\cite{meng2021persistent}. This performance advantage is visually detailed in Figure~\ref{fig:PCC}a, with the linear correlation illustrated in Figure~\ref{fig:PCC}b.

\subsection{Robustness and Generalization in Affinity Prediction}
While CASF benchmarks provide a standard metric for scoring power, practical drug discovery often involves evaluating targets with low sequence similarity to known data. To evaluate model generalization under such conditions, we conducted experiments on the LBA30 and LBA60 datasets \cite{townshend2020atom3d}, where training and test proteins are separated under strict sequence-identity thresholds. This partitioning strategy evaluates the model’s true generalization ability across unseen protein families.

We compared TopoLIA with 12 representative baseline models spanning sequence-based, structure-based, and pre-trained approaches. As shown in Figure~\ref{fig:PCC}d, Figure~\ref{fig:PCC}e, and Table S6, TopoLIA achieves the best overall performance across both sequence-identity splits. On the highly challenging LBA30 dataset, TopoLIA secures a leading PCC of 0.641 and an RMSE of 1.307 $pK_d$, outperforming recent structure-based methods like BindNet~\cite{li2025harnessing}. On LBA60, TopoLIA significantly outperforms the current state-of-the-art method SableBind, achieving a PCC of 0.806 and lowering the RMSE to 1.220 $pK_d$. These results demonstrate that the topological signatures learned by TopoLIA support robust affinity prediction and enable effective generalization across diverse protein families.

Beyond sequence identity evaluations, we further assessed the robustness and performance consistency of TopoLIA using 10-fold cross-validation on the PDBbind v.2020 refined set~\cite{liu2015pdb}. As one of the most comprehensive and up-to-date collections of high-quality protein--ligand structures, this dataset poses a substantial predictive challenge due to its expanded structural diversity. The cross-validation setting mitigates sampling biases and ensures a robust evaluation across diverse data partitions. TopoLIA achieved a PCC of 0.811 and an RMSE of 1.152 kcal/mol (Figure~\ref{fig:PCC}g). In comparison, recent graph-based models such as PLAIG~\cite{samudrala2025plaig} and GNNSeq~\cite{dandibhotla2025gnnseq} yielded considerably lower PCC scores of 0.763 and 0.784, accompanied by RMSEs of 1.251 kcal/mol and 1.336 kcal/mol, respectively. These results confirm the robustness of TopoLIA on large and structurally diverse protein--ligand datasets.

\subsection{Docking Power Evaluation}  
The capacity of a scoring function (SF) to distinguish native binding poses from non-native decoys is a critical metric for molecular docking. We evaluated the docking proficiency of TopoLIA on the CASF-2013 benchmark using the standard protocol where a prediction is considered successful if the pose with the highest predicted binding affinity falls within a 2.0~\AA{} root-mean-square deviation (RMSD) threshold of the experimentally determined native structure. To ensure statistical robustness, the final success rates are derived from the median outputs of a multi-model ensemble, including ten independent TopoLIA runs, forty GBT models utilizing Euclidean distance-based persistent spectra features, and twenty GBT models utilizing electrostatic distance-based persistent spectra features.

Recent deep learning models have improved docking performance~\cite{gentile2021automated}, with representative methods including DeepDock (62.1\%)~\cite{mendez2021geometric}, OnionNet-SFCT (76.8\%)~\cite{zheng2022improving}, and RTMScore (80.7\%)~\cite{shen2022boosting}. On CASF-2013, TopoLIA achieved a superior docking success rate of 88\%. As comprehensively illustrated in Figure~\ref{fig:PCC}c (with the full benchmark hierarchy provided in Table S5), this performance significantly outperforms a wide array of established classical scoring functions, including AutoDock Vina (85\%) and $\Delta_{\text{vina}}\text{RF}_{20}$ (87\%).

Developing SFs that perform well in both scoring and docking power remains challenging in computational drug discovery. Many established docking packages provide reliable structural pose predictions but limited binding affinity estimation. For instance, on the CASF-2013 benchmark, the ChemPLP@GOLD SF achieves a docking success rate of 81\% but an unsatisfactory PCC of 0.579 for binding affinity prediction. Similarly, AutoDock Vina achieves an 85\% docking success rate and a PCC of 0.564. In contrast, TopoLIA demonstrates robust versatility across both tasks, achieving a PCC of 0.814 for binding affinity prediction and an 88\% docking success rate on the same dataset.

\begin{figure}[htp] 
\centering  
\includegraphics[width=1.0\linewidth]{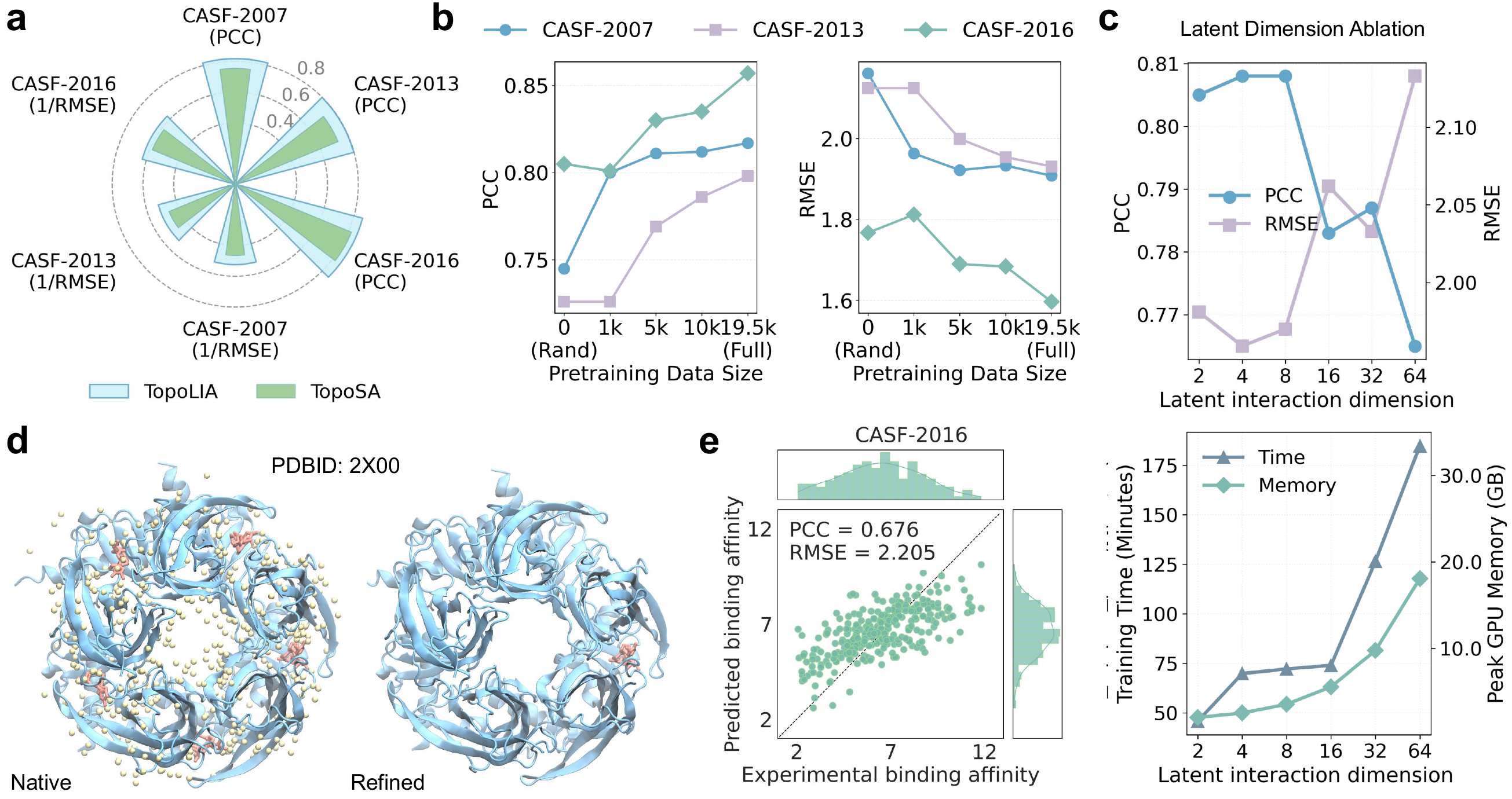} 
\caption{Comprehensive performance evaluation, scaling behavior, ablation analysis, and structural insights of the TopoLIA framework. 
(a): Radial bar chart comparing the predictive performance of TopoLIA, equipped with latent interaction attention, and its standard-attention counterpart, TopoSA. Metrics include PCC and inverse RMSE across three benchmarks.
(b): Downstream performance scaling as a function of pretraining data volume. The line plots illustrate the impact of pretraining with increasing amounts of unlabeled data, ranging from random initialization without pretraining (`Rand', 0) to the complete dataset (`Full', 19.5k), demonstrating consistent monotonic improvements in both PCC (left) and RMSE (right) across all three benchmarks.
(c): Latent interaction space dimension ablation. The top panel evaluates the predictive performance on CASF-2007 as a function of the latent rank $r$. The bottom panel details the corresponding computational scalability, tracking the cumulative training time and peak GPU memory allocation.
(d): Comparative structural analysis for a predictive outlier (PDBID: 2X00), contrasting the native protein-ligand complex (left) with the refined representation (right) utilized by the TopoLIA model.
(e): Linear probing performance on CASF-2016. The joint plot displays the correlation between experimental and predicted binding affinities when the pretrained TopoLIA encoder is entirely frozen and only a minimalist linear evaluation head is trained.}
\label{fig:ablation_combined} 
\end{figure}

\subsection{Ablation Study on Latent Interaction Attention}

To isolate the contribution of latent interaction attention (LIA), we constructed a baseline model, denoted TopoSA, that retains the same persistent spectral representations and overall transformer architecture as TopoLIA, while replacing LIA with standard self-attention \cite{vaswani2017attention}. This controlled comparison allows us to directly assess whether introducing source-target-dependent feature transformation beyond the scalar weighting of standard attention improves protein--ligand interaction modeling.

We compared TopoLIA with TopoSA on the CASF-2007, CASF-2013, and CASF-2016 benchmarks, as shown in the radial bar chart of Figure \ref{fig:ablation_combined}a. Ten independent runs with different random seeds were performed and the final PCC and RMSE were computed from the averaged predictions of the ten models. To provide a unified geometric representation in polar coordinates, we report PCC together with inverse RMSE, such that outward radial expansion consistently indicates improved predictive performance. Across all evaluation axes, the TopoLIA profile (light blue) consistently extends beyond that of the TopoSA baseline (dark green). Quantitatively, TopoLIA achieves consistent improvements across all three benchmarks. Specifically, TopoLIA improves PCC from 0.754 to 0.817 and reduces RMSE from 2.153 kcal/mol to 1.908 kcal/mol on CASF-2007, improves PCC from 0.724 to 0.798 and reduces RMSE from 2.144 kcal/mol to 1.931 kcal/mol on CASF-2013, and improves PCC from 0.819 to 0.857 and reduces RMSE from 1.712 kcal/mol to 1.597 kcal/mol on CASF-2016. These consistent gains across all three benchmarks demonstrate the effectiveness of latent interaction attention relative to standard attention.

Because the two models share the same persistent spectral representations and surrounding transformer architecture, the observed performance differences isolate the contribution of the attention mechanism. While standard self-attention effectively identifies relevant source information through scalar attention weights, it does not explicitly adapt the transformation of that information to different targets. The consistent advantage of TopoLIA therefore demonstrates the benefit of source-target-dependent feature transformation for modeling heterogeneous protein--ligand interactions, supporting the central design principle of LIA that effective information exchange involves not only determining how much information to communicate, but also how that information should be transformed.

\subsection{Ablation Study on the Dimensionality of the Latent Interaction Space}

To investigate how the capacity of the latent interaction space affects source-target information transformation in LIA, we systematically varied the dimension of the latent interaction space over $r \in \{2, 4, 8, 16, 32, 64\}$. Ten independent training runs with different random seeds were performed for each latent dimension, and the final PCC and RMSE were computed from the averaged predictions of the ten models. This experiment directly tests our hypothesis that effective protein--ligand information exchange can be mediated through a substantially lower-dimensional interaction space than the full feature space.

The results on CASF-2007 reveal a clear dependence of predictive performance on the dimensionality of the latent interaction space. As shown in the top panel of Figure~\ref{fig:ablation_combined}c, TopoLIA achieves its strongest generalization within a compact range of $r\in[4,8]$, with the complete numerical results and computational costs summarized in Table S9. At $r=4$, the model achieves a PCC of 0.808 and an RMSE of 1.959 kcal/mol, while $r=8$ yields a nearly identical PCC of 0.808 and an RMSE of 1.970 kcal/mol. Beyond this range, increasing the latent dimension does not improve predictive performance. Instead, performance progressively declines, reaching a PCC of 0.765 and an RMSE of 2.133 kcal/mol at $r=64$. These results indicate that a larger interaction space is not necessarily more expressive in a useful sense and that protein--ligand interactions can be effectively represented through a relatively low-dimensional latent space.

The computational analysis further demonstrates an additional practical advantage of maintaining a compact latent interaction space. As shown in the bottom panel of Figure~\ref{fig:ablation_combined}c, computational cost increases with the latent dimension, and an Out-of-Memory failure occurs at $r=111$ on a 32 GB NVIDIA Tesla V100S GPU under the experimental configuration used here. Thus, increasing the dimensionality of the interaction space not only fails to improve predictive accuracy beyond the optimal range, but also introduces additional computational and memory demands.

Taken together, these results provide empirical support for the low-dimensional latent interaction hypothesis underlying LIA. The optimal performance at intermediate dimensions suggests that effective source-target biomolecular information exchange does not require transformations in the full feature space. Instead, a compact latent interaction space appears sufficient to capture the interaction-relevant information exchanged between molecular features. Conversely, excessively large latent dimensions introduce additional degrees of freedom without corresponding predictive gains and may reduce generalization. The existence of an optimal intermediate dimension therefore suggests that the dimensionality of the latent interaction space is not merely a computational hyperparameter, but an important architectural factor governing how effectively LIA models protein--ligand interactions.

\subsection{Ablation Study on the Impact of Pretraining}
To assess the efficacy of self-supervised pretraining, we performed a systematic ablation analysis evaluating the downstream performance across a spectrum of pretraining data scales. The latent interaction transformer encoder was pretrained using progressively larger subsets of unlabeled protein--ligand complexes and then fine-tuned on downstream benchmarks. This analysis evaluates both the empirical utility of pretraining and the dependence of downstream performance on pretraining data scale. The hyperparameters utilized for pretraining and fine-tuning are detailed in Table S2.

Empirical results across three benchmarks demonstrate that pretraining consistently improved downstream performance. As shown in Figure~\ref{fig:ablation_combined}b, we varied the number of the unlabeled pretraining samples, from 0 (random initialization) to 1k, 5k, 10k, and the complete set of ~19.5k samples, with full quantitative results summarized in Table S7. The resulting trajectories reveal a clear law: expanding the pretraining data size improves downstream performance. Notably, using 5,000 unlabeled complexes already leads to a substantial improvement over random initialization, while further increasing the data size continues to enhance performance without obvious saturation. On CASF-2016, leveraging the full pretraining regime elevated the PCC from 0.805 to 0.857, while simultaneously reducing the RMSE from 1.767 to 1.597 kcal/mol. Figure S2 visually corroborates this advantage, demonstrating a markedly tighter linear fit for the pretrained model compared to the random baseline.

To further validate the expressive power of the learned topological representations, we evaluated a linear probing configuration. By completely freezing the pretrained encoder and exclusively training a minimalist linear projection head (with a fine-tuning learning rate of $8 \times 10^{-4}$), the model still delivered strong predictive performance. It achieved a PCC of 0.613 (RMSE: 2.555 kcal/mol) on CASF-2007 and a PCC of 0.650 (RMSE: 2.356 kcal/mol) on CASF-2013. Particularly on the CASF-2016 benchmark, the model achieved a PCC of 0.676 (RMSE: 2.205 kcal/mol), with the robust linear correlation between experimental and predicted binding affinities explicitly visualized in Figure~\ref{fig:ablation_combined}e. These results provide concrete evidence that our self-supervised pretraining phase successfully maps fundamental geometric and topological inputs into rich, highly informative structural representations, yielding robust predictive utility even in the absence of deep, task-specific adaptation.

\section{Discussion}
In this study, we introduced TopoLIA, centered on a new latent interaction attention (LIA) mechanism for modeling heterogeneous biomolecular information exchange. Conventional attention primarily determines how much information should be exchanged through scalar attention weights, while the transformation of a source value is shared across different targets. LIA extends this paradigm by additionally learning how source information should be transformed for a particular target. We integrate this interaction-dependent attention mechanism with multiscale persistent spectral representations of three-dimensional protein--ligand complexes, resulting in a framework in which topological structure informs both molecular representation and neural information exchange.

The comparison with standard self-attention provides direct evidence for the contribution of LIA. Under the same configurations, including the same persistent spectral representations and surrounding transformer architecture, replacing LIA with standard self-attention consistently reduces predictive performance across the evaluated protein--ligand benchmarks. This controlled comparison isolates the contribution of the attention mechanism and supports the central design principle of LIA: effective information exchange involves not only determining how much information should be communicated, but also how that information should be transformed for a particular interaction.

The latent-dimension analysis further reveals an important property of this interaction mechanism. TopoLIA achieves its strongest generalization within a compact intermediate latent dimension, with optimal performance observed around $r=4$--$8$, whereas increasing the interaction dimension beyond this range provides no additional predictive benefit and instead reduces generalization performance. These results provide empirical support for our hypothesis that interaction-relevant information can be effectively exchanged through a substantially lower-dimensional space than the full feature space. Importantly, this does not imply that protein--ligand interactions themselves are intrinsically low-dimensional. Rather, it suggests that effective source-target communication within LIA can be mediated through a compact latent interaction space. The existence of an optimal compact dimension further suggests that the dimensionality of the latent interaction space plays an important role in governing the effectiveness of LIA, rather than serving merely as a computational design choice. The associated reductions in computational and memory costs provide an additional practical advantage.

Persistent spectral representation and self-supervised pretraining provide complementary contributions at the structural representation level. The persistent spectral embeddings encode multiscale topological and geometric characteristics of protein--ligand complexes and serve as structural tokens for masked self-supervised learning. By reconstructing masked spectral tokens from their molecular context, the model learns transferable multiscale structural dependencies before exposure to downstream task labels. The consistent improvement in downstream performance with increasing pretraining scale, together with successful transfer across scoring and docking tasks, demonstrates that TopoLIA benefits from larger collections of unlabeled molecular structures. These results further indicate that persistent spectral tokens provide transferable structural information and suggest that continued scaling to larger collections of experimentally determined and computationally predicted structures may further improve generalization.

The framework also suggests several directions for extension. Because LIA is not specific to protein--ligand systems, the same interaction-dependent transformation principle may be applicable to other heterogeneous biomolecular interactions, including protein--protein, protein--peptide, protein--nucleic acid, and mutation-induced interaction modeling. At the representation level, richer mathematical constructions, including persistent path Laplacians \cite{wang2023persistent}, persistent hyperdigraph Laplacians \cite{chen2023persistent}, and persistent Dirac operators \cite{ameneyro2024quantum,suwayyid2024persistent}, may provide additional sensitivity to directed, multi-body, and higher-order molecular interactions. At the same time, our error analysis in Figure \ref{fig:ablation_combined}d and Section S4.2 highlights an important limitation of the current molecular representation. The present formulation primarily models the protein and ligand and therefore captures environmental factors only indirectly. Complexes whose binding behavior is strongly influenced by structural water molecules, metal ions, cofactors, or other components of the local chemical environment appear to remain more challenging. Explicitly incorporating such environmental components into the molecular representation and topological filtration process may improve both biological fidelity and predictive accuracy.

Overall, TopoLIA suggests that topology-inspired learning can extend beyond structural representation to the design of neural information exchange. By combining multiscale persistent spectral representations with interaction-dependent feature transformations, TopoLIA demonstrates how mathematical structure can inform both what molecular information is represented and how that information is exchanged within a neural network. More broadly, LIA provides a principled mechanism for modeling heterogeneous interactions through source-target-dependent feature transformation and can be directly applied to all other large language models (LLMs) involving heterogeneous components.

\section{Methods}
\subsection{Latent Interaction Attention}
Standard attention determines the contribution of a source token to a target token through a scalar attention coefficient. However, once the source value is computed, the same transformed value is shared across different targets. To additionally model how information should be transformed for a particular source-target interaction, we introduce latent interaction attention (LIA). LIA retains the query-key attention mechanism for determining how much information is exchanged, while introducing an interaction-dependent feature transformation through a latent interaction space.

Let $h_x^{(\ell)} \in \mathbb{R}^{d}$ denote the representation of cell $x$ at layer $\ell$, where $d$ is the feature dimension. For clarity, we describe a single attention head and the multi-head formulation applies the same mechanism independently across heads. Given learnable projection matrices $W_q,W_k,W_v\in\mathbb{R}^{d\times d}$, the query, key, and value representations are defined as
\begin{equation}
q_x=W_qh_x^{(\ell)},\qquad
k_y=W_kh_y^{(\ell)},\qquad
v_y=W_vh_y^{(\ell)}.
\end{equation}
As in standard scaled dot-product attention, the scalar attention coefficient from source $y$ to target $x$ is
\begin{equation}
a_{y\to x}=
\frac{
\exp\left(\langle q_x,k_y\rangle/\sqrt{p}\right)
}{
\sum_{y'\in\mathcal{N}(x)}
\exp\left(\langle q_x,k_{y'}\rangle/\sqrt{p}\right)
},
\label{eq:lia_attention}
\end{equation}
where $\mathcal{N}(x)$ denotes the set of source cells (neighbors) attended to by $x$. The coefficient $a_{y\to x}$ therefore determines how much information is transferred from $y$ to $x$.

LIA further introduces a source-dependent projection and a target-dependent projection to determine how this information is transformed. For each source $y$ and target $x$, feature-dependent projection networks generate
\begin{equation}
e_y^\top \in \mathbb{R}^{r\times d},
\qquad
d_x \in \mathbb{R}^{d\times r},
\end{equation}
where $r$ denotes the dimension of the latent interaction space. Specifically,
\begin{equation}
e_y=\phi_e\left(h_y^{(\ell)}\right),
\qquad
d_x=\phi_d\left(h_x^{(\ell)}\right),
\end{equation}
with the outputs reshaped into the corresponding projection matrices. The source value is first projected into the latent interaction space through $e_y^\top$,
\begin{equation}
z_y=e_y^\top v_y\in\mathbb{R}^{r},
\end{equation}
and is subsequently mapped back to the feature space through the target-dependent projection $d_x$. The resulting message received by target $x$ is therefore
\begin{equation}
m_x=
\sum_{y\in\mathcal{N}(x)}
a_{y\to x}
d_xe_y^\top v_y.
\label{eq:lia_message}
\end{equation}

This formulation separates two complementary components of information exchange. The scalar coefficient $a_{y\to x}$, obtained from query-key similarity, determines \emph{how much} information source $y$ contributes to target $x$. In parallel, the composition
\begin{equation}
T_{y\to x}=d_xe_y^\top
\end{equation}
defines a source-target-dependent transformation that determines \emph{how} the source information is represented for that target. Importantly, LIA does not explicitly parameterize an independent $d\times d$ transformation matrix for every source-target pair. Instead, the interaction transformation is mediated through the shared $r$-dimensional latent interaction space: the source-dependent projection $e_y^\top$ encodes source information into this space, whereas the target-dependent projection $d_x$ decodes the resulting latent representation according to the target context. Thus, different targets can receive differently transformed versions of the same source information even when they attend to the same source token.

As illustrated in Figure~\ref{fig:workflow}d, LIA follows the computation in Eq.~\ref{eq:lia_message}. For each source-target pair $(y,x)$, the source value $v_y$ is first encoded into the latent interaction space through the source-dependent projection $e_y^\top$ and then decoded back into the feature space through the target-dependent projection $d_x$, yielding the interaction-dependent transformation $d_xe_y^\top v_y$. The transformed source information is subsequently weighted by the scalar attention coefficient $a_{y\to x}$, obtained from query-key similarity, and aggregated over all source tokens to produce $m_x$. The output $m_x$ is then incorporated into the standard transformer residual, normalization, and feed-forward operations. Replacing conventional self-attention with LIA while retaining the remaining transformer structure defines the latent interaction transformer (LIT) layer.

The latent formulation also avoids explicitly constructing dense pair-specific transformations. An unrestricted source--target transformation in $\mathbb{R}^{d\times d}$ would require $\mathcal{O}(d^2)$ operations for each interacting pair, resulting in an application cost of $\mathcal{O}(md^2)$ for $m$ interactions. Although LIA is conceptually expressed as
$$m_x=\sum_{y\in\mathcal{N}(x)}a_{y\to x}d_xe_y^\top v_y,$$
the target-dependent projection $d_x$ is independent of the source index $y$ and can therefore be applied after aggregation:
$$m_x=d_x\sum_{y\in\mathcal{N}(x)}a_{y\to x}e_y^\top v_y.$$
This equivalent formulation enables the source projection $e_y^\top v_y$ to be computed once for each source token, followed by attention-weighted aggregation directly in the $r$-dimensional latent interaction space and a single target projection for each target token, as summarized in Algorithm \ref{alg:LIA}. For $n$ tokens and $m$ source-target interactions, these operations require $\mathcal{O}(ndr)$ for the source projections, $\mathcal{O}(mr)$ for latent-space aggregation, and $\mathcal{O}(ndr)$ for the target projections, yielding an overall interaction-application complexity of $\mathcal{O}(ndr+mr)$. This estimate excludes the shared neural networks used to generate $e_y$ and $d_x$, as well as the standard query, key, and value projections. When $r\ll d$, LIA therefore provides a compact implementation of source-target-dependent feature transformation without explicitly applying a dense $d\times d$ transformation to every interacting pair.

\begin{algorithm}[H]
\caption{Latent Interaction Attention}
\label{alg:LIA}
\begin{algorithmic}[1]
    \State \textbf{Input:} Node representations $h^{(\ell)}_x \in \mathbb{R}^d$, Topological neighborhood $\mathcal{N}$
    \State \textbf{Output:} Updated node representations $h^{(\ell+1)}_x$
    \Statex
    \For{each $x \in X$} \Comment{ Feature projection}
        \State $q_x, k_x, v_x \gets W_q h^{(\ell)}_x, W_k h^{(\ell)}_x, W_v h^{(\ell)}_x$ \Comment{Head-wise projections}
        \State $e_x \gets \phi_e(h^{(\ell)}_x) \in \mathbb{R}^{d \times r}$ \Comment{Source-dependent latent projection}
        \State $d_x \gets \phi_d(h^{(\ell)}_x) \in \mathbb{R}^{d \times r}$ \Comment{Target-dependent latent projection }
    \EndFor
    \Statex
    \For{each $x \in X$} \Comment{ Aggregate information for target $x  $}
        \State $z_x^{\text{mix}} \gets \mathbf{0} \in \mathbb{R}^r$ \Comment{Initialize latent interaction representation}
        \For{each $y \in \mathcal{N}(x)$} \Comment{Inner loop over topological neighbors}
            \State $a_{y\to x} \gets \text{Softmax}(\frac{q_x^\top k_y}{\sqrt{d_{k}}})$ \Comment{Compute attention score (Eq. \ref{eq:attention})}
            \State $z_y \gets e_y^\top v_y \in \mathbb{R}^r$ \Comment{Encode source value to latent interaction space}
            \State $z_x^{\text{mix}} \gets z_x^{\text{mix}} + a_{y\to x} z_y$ \Comment{Aggregate attention-weighted latent information}
        \EndFor
        \State $m_x \gets d_x z_x^{\text{mix}} \in \mathbb{R}^d$ \Comment{Decode latent information for target $x$}
        \State $h^{(\ell+1)}_x \gets \beta(h^{(\ell)}_x, m_x)$ \Comment{Apply update function}
    \EndFor
    \State \Return $\{h^{(\ell+1)}_x\}$
% \EndProcedure
\end{algorithmic}
\end{algorithm}

\subsection{Topological Persistent Spectral Theory}
To characterize the intricate and multiscale nature of protein--ligand interactions, we employ a persistent spectral framework \cite{wang2020persistent} capable of capturing both topological invariants and geometric variations. The mathematical foundation begins with an abstract simplicial complex $\mathcal{K}$ defined over a vertex set $\mathcal{V}$, representing atomic coordinates. $\mathcal{K}$ is a collection of non-empty subsets of $\mathcal{V}$ such that for every simplex $\tau \in \mathcal{K}$, all its non-empty subsets are also contained in $\mathcal{K}$. An element $\tau \in \mathcal{K}$ consisting of $k+1$ vertices is called a $k$-simplex. The dimension of a $k$-simplex is defined as $k$, and the dimension of a simplicial complex is defined as the maximal dimension of its simplices. Any graph can be regarded as a one-dimensional simplicial complex.
\begin{figure}[ht] 
\centering  
\includegraphics[width=0.9\linewidth]{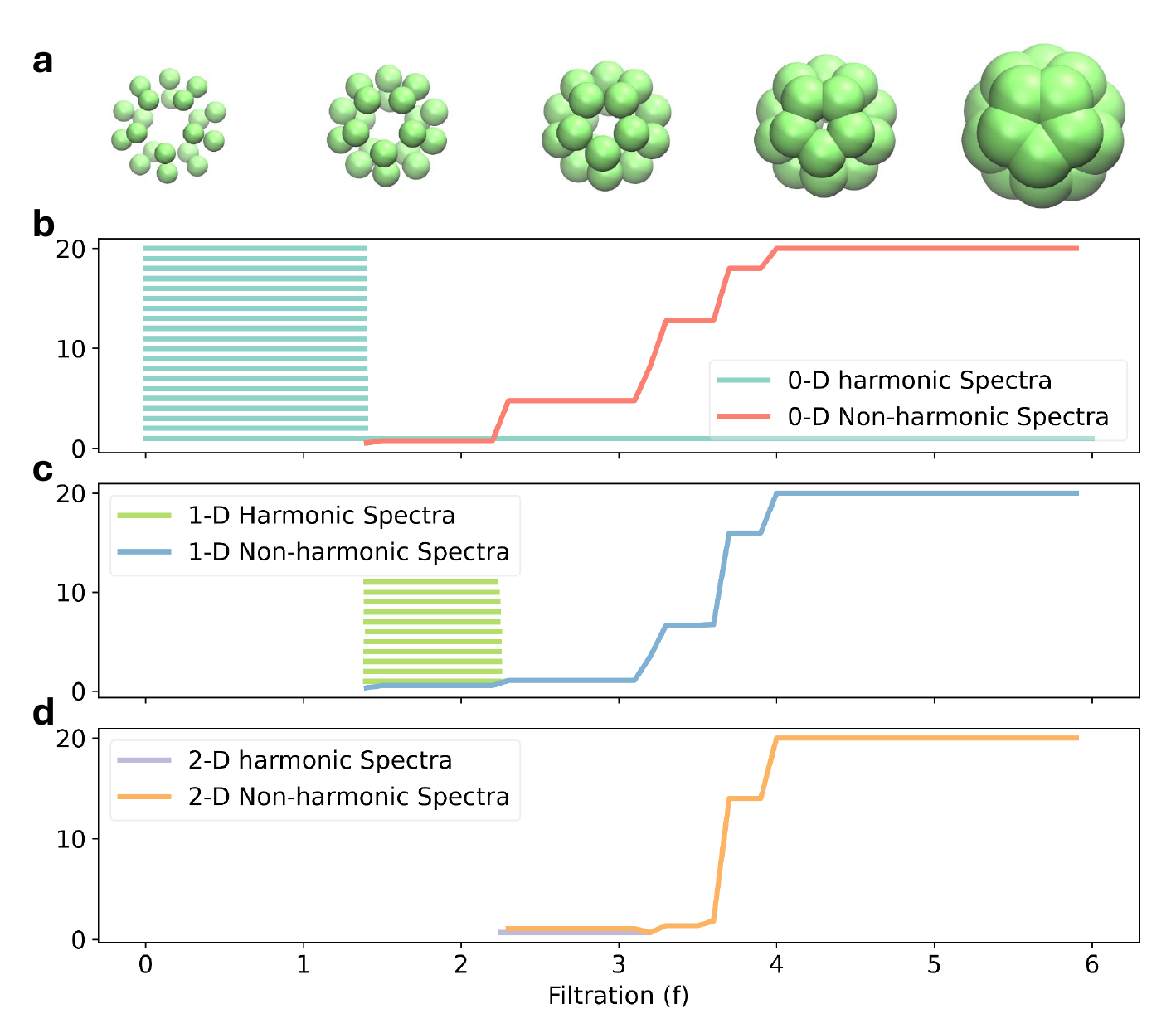} 
\caption{Multiscale characterization of persistent Laplacian spectra. (a) Vietoris--Rips filtration of the $C_{20}$ structure. (b) Zero-dimensional persistent Laplacian spectra. (c) One-dimensional persistent Laplacian spectra. (d) Two-dimensional persistent Laplacian spectra. The harmonic spectra recover persistent homology information, while the non-harmonic spectra capture additional geometric information.}
\label{fig:spectral_characterization}
\end{figure}
An oriented simplicial complex $\mathcal{K}$ is a simplicial complex equipped with an ordering of its vertex set. Each simplex of $\mathcal{K}$ inherits an orientation from the vertex ordering; a simplex together with this induced order is referred to as an oriented simplex. For simplicity, we refer to oriented simplices simply as simplices throughout the rest of this exposition. We define the $k$-th chain group $\mathcal{C}_k(\mathcal{K})$ as the vector space over a field $\mathbb{F}$ with a basis formed by $k$-simplices. The structural relationships within the complex are governed by the boundary operator $\mathcal{B}_k : \mathcal{C}_k(\mathcal{K}) \to \mathcal{C}_{k-1}(\mathcal{K})$, defined on a $k$-simplex $\tau = [v_0, \dots, v_k]$ as:
\begin{equation}
    \mathcal{B}_k(\tau) = \sum_{j=0}^{k} (-1)^j [v_0, \dots, v_{j-1}, v_{j+1}, \dots, v_k].
\end{equation}
These operators satisfy the fundamental property $\mathcal{B}_k \circ \mathcal{B}_{k+1} = 0$, forming a chain complex:
\begin{equation}
    \cdots \xrightarrow{\mathcal{B}_{k+1}} \mathcal{C}_k(\mathcal{K}) \xrightarrow{\mathcal{B}_k} \mathcal{C}_{k-1}(\mathcal{K}) \xrightarrow{\mathcal{B}_{k-1}} \cdots \xrightarrow{\mathcal{B}_1} \mathcal{C}_0(\mathcal{K}) \to 0.
\end{equation}
The $k$-th Betti number, denoted $\beta_k$, is the rank of the homology group $\mathcal{H}_k(\mathcal{K}) = \text{ker}(\mathcal{B}_k) / \text{im}(\mathcal{B}_{k+1})$. Intuitively, $\beta_0$ counts the number of connected components, $\beta_1$ the number of loops, and $\beta_2$ the number of voids. These topological invariants capture essential structural information about the simplicial complex.

To extract persistent spectral information, we consider the standard inner product on the chain groups, where for any two simplices $\tau$ and $\upsilon$, $\langle \tau, \upsilon \rangle = \delta_{\tau,\upsilon}$ with $\delta_{\tau,\upsilon}$ being the Kronecker delta. Let $\mathcal{B}_k^*$ denote the adjoint of $\mathcal{B}_k$ under this inner product. The $k$-th combinatorial Laplacian operator is then defined as:
\begin{equation}
    \mathcal{L}_k = \mathcal{B}_k^* \mathcal{B}_k + \mathcal{B}_{k+1} \mathcal{B}_{k+1}^*.
\end{equation}
In matrix form, $\mathcal{B}_k^*$ is the transpose of $\mathcal{B}_k$. A fundamental property of the Laplacian is that the nullity of $\mathcal{L}_k$ equals the $k$-th Betti number $\beta_k$. For protein--ligand systems, we consider a filtration, a nested sequence of simplicial complexes: $\mathcal{K}_0 \subseteq \mathcal{K}_1 \subseteq \dots \subseteq \mathcal{K}_n$. Applying the homology functor to this sequence yields a sequence of homology groups:
\begin{equation}
    \mathcal{H}_k(\mathcal{K}_0) \xrightarrow{i_0} \mathcal{H}_k(\mathcal{K}_1) \xrightarrow{i_1} \cdots \xrightarrow{i_{n-1}} \mathcal{H}_k(\mathcal{K}_n),
\end{equation}
where $i_j$ is the homomorphism induced by the inclusion map from $\mathcal{K}_j$ to $\mathcal{K}_{j+1}$. Let $i_{s,t}$ denote the composition $\mathcal{H}_k(\mathcal{K}_s) \to \mathcal{H}_k(\mathcal{K}_t)$. The $k$-th persistent homology from $\mathcal{K}_s$ to $\mathcal{K}_t$ is defined as $\mathcal{H}_k^{s,t} = \text{Im}(i_{s,t})$, which represents the homology classes that appear at $\mathcal{K}_s$ and remain at $\mathcal{K}_t$.

Let $\mathcal{C}_k^s = \mathcal{C}_k(\mathcal{K}_s)$. Consider the subspace $\mathcal{C}_k^{s,t} = \{ \alpha \in \mathcal{C}_k(\mathcal{K}_t) \mid \mathcal{B}_k(\alpha) \in \mathcal{C}_{k-1}(\mathcal{K}_s) \}$. Then we have:
\begin{equation}
    \mathcal{C}_{k+1}^{s,t} \xrightarrow{\mathcal{B}_{k+1}^{s,t}} \mathcal{C}_k^s \xrightarrow{\mathcal{B}_k} \mathcal{C}_{k-1}^s,
\end{equation}
where $\mathcal{B}_{k+1}^{s,t}$ is the restriction of the boundary operator on the subspace $\mathcal{C}_{k+1}^{s,t}$. The $k$-th persistent Laplacian from $\mathcal{K}_s$ to $\mathcal{K}_t$ is defined as:
\begin{equation}
    \mathcal{L}_k^{s,t} = \mathcal{B}_k^* \mathcal{B}_k + \mathcal{B}_{k+1}^{s,t} (\mathcal{B}_{k+1}^{s,t})^*.
\end{equation}
While the zero eigenvalues (harmonic spectra) represent persistent topological invariants through the relation $\text{rank}(\mathcal{H}_k^{s,t}) = \text{nullity}(\mathcal{L}_k^{s,t})$, the non-zero eigenvalues (non-harmonic spectra) provide essential geometric information regarding the connectivity and density of the molecular interface. These combined features form the structured topological sequences that are processed by the latent interaction transformer for high-fidelity binding affinity prediction.

To ground this mathematical framework, Figure \ref{fig:spectral_characterization} illustrates the multiscale spectral pipeline using a $C_{20}$ point cloud. 
(a) shows its Vietoris–Rips filtration, while (b)–(d) present the corresponding 0-, 1-, and 2-dimensional persistent Laplacian spectra. The harmonic components in (b)–(d) recover persistent homology, capturing connected components, pentagonal loops, and the enclosed void, respectively. In contrast, the non-harmonic eigenvalue curves in (b)–(d) vary with the addition of edges, triangles, and higher-dimensional simplices, thereby encoding geometric changes even after the persistent homology has stabilized. Together, these results show that persistent Laplacian spectra retain multiscale topological information while providing additional geometric information beyond persistent homology.

\subsection{Multiscale Topological  Spectral Characterization}
As demonstrated by the empirical evaluations in Section S4.1, capturing a holistic representation of the protein--ligand interface requires the synergistic integration of both structural shapes and charge-based environments. To mathematically formalize the geometric--electrostatic dual topological deep learning  (GE-TopoLIA) ensemble, we detail the construction of its two specialized constituent modules: G-TopoLIA and E-Phys.

To achieve a granular and physically meaningful representation of the chemical environment, both modules employ an element-specific (ES) topological model. This approach partitions the molecular complex into distinct atom sets based on their respective atomic species, allowing the network to distinguish between varying chemical identities. Specifically, the protein structure is decomposed into four primary atom sets representing Carbon (C), Nitrogen (N), Oxygen (O), and Sulfur (S). The ligand structure is similarly partitioned into five sets, encompassing Carbon (C), Nitrogen (N), Oxygen (O), Sulfur (S), and Phosphorus (P), resulting in $20=4 \times 5$ potential element-specific interaction combinations for a given complex. Let $\mathcal{E}_{\text{p}}$ and $\mathcal{E}_{\text{l}}$ denote the sets of 3D Cartesian coordinates for these specific atom types in the protein and ligand, respectively. By focusing exclusively on the bipartite interactions between a set from $\mathcal{E}_{\text{p}}$ and a set from $\mathcal{E}_{\text{l}}$, the model isolates the critical cross-boundary relationships that govern binding affinity. This element-specific decomposition results in a series of distinct bipartite graphs, ensuring that the subsequent persistent spectral analysis captures the unique physical and chemical contributions of each atomic pairing.

The primary structural backbone of our framework, the G-TopoLIA module, focuses on capturing the intrinsic geometric shape and complementarity of the binding pocket. Spatial relationships at the interface are formally defined by an interactive Euclidean distance matrix $D$. To ensure the model strictly emphasizes the protein--ligand interface, we assign an infinite value to distances between atoms within the same molecule, effectively ignoring protein--protein or ligand--ligand interactions. We define the binding core region by considering protein atoms within 10~\AA{} of any ligand atom. The interactive Euclidean distance matrix is formulated as follows:
\begin{equation}
    D(i, j) = 
    \begin{cases} 
    \| r_i - r_j \| & \text{if } (r_i \in \mathcal{E}_{\text{p}} \land r_j \in \mathcal{E}_{\text{l}}) \lor (r_i \in \mathcal{E}_{\text{l}} \land r_j \in \mathcal{E}_{\text{p}}) \\
    \infty & \text{otherwise}
    \end{cases}
\end{equation}
where $r_i$ and $r_j$ denote the coordinates of the $i$-th and $j$-th atoms, and $\| \cdot \|$ represents the standard Euclidean distance. This distance matrix serves as the foundation for the subsequent Vietoris-Rips construction, allowing the G-TopoLIA to extract multiscale geometric signatures through a latent interaction transformer architecture.
% \subsubsection{Electrostatic Interaction and Charge-Modulated Filtration}

To complement these geometric features with physicochemical information, the E-Phys module models the electrostatic energy landscape through an interactive matrix $D_{\text{elec}}$. Given the fundamental importance of hydrogen atoms in electrostatic interactions, H atoms are explicitly included in this module, leading to a total of $30=5 \times 6$ distinct atom combinations. We utilize the PDB2PQR software \cite{dolinsky2007pdb2pqr} to generate partial charges for protein atoms, while ligand partial charges are retrieved from the PDBbind database. The electrostatic distance function is defined as:
\begin{equation}
    D_{\text{elec}}(i, j) = 
    \begin{cases} 
    1 + \frac{1}{1 + \exp\left(- \frac{c q_i q_j}{\| r_i - r_j \|} \right)} & \text{if } (r_i \in \mathcal{E}_{\text{p}} \land r_j \in \mathcal{E}_{\text{l}}) \lor (r_i \in \mathcal{E}_{\text{l}} \land r_j \in \mathcal{E}_{\text{p}}) \\
    \infty & \text{otherwise}
    \end{cases}
\end{equation}
In this formulation, $q_i$ and $q_j$ represent the atomic partial charges, and $c$ is a constant scaling parameter, usually taken as 100. Similar to the Euclidean approach, electrostatic interactions within either the protein or ligand are dismissed by setting their values to $\infty$. This modulation of the topological filtration by atomic charges allows the E-Phys module to capture the electrostatic nuances of the binding environment.

We generate sequences of simplicial complexes using a Vietoris-Rips construction across a filtration range of 0~\AA{} to 10~\AA{} with a step size of 0.1~\AA{} for the Euclidean distance and a filtration of 0 to 1 with step 0.02 for the electrostatic distance. We focus exclusively on 0-D persistent spectral information to characterize the interface. For the G-TopoLIA module, the spectral features are composed of both harmonic and non-harmonic components. The harmonic spectra are computed by determining the multiplicity of zero eigenvalues, while the non-harmonic spectra are obtained through seven statistical measures: the maximum, minimum, mean, sum, standard deviation, variance, and the count of non-zero eigenvalues. This results in a feature tensor of size $20 \times 100 \times 8$, which is processed by the latent interaction transformer.

For the E-Phys module, we characterize the spectral moments of the simplicial complex using Riemann Zeta functions, $\zeta(s)$, as introduced in \cite{liu2022dowker}. For a set of eigenvalues $\{\lambda_1, \lambda_2, \dots, \lambda_n\}$, the spectral moments $\zeta(s)=\sum_{i=1}^n\frac{1}{\lambda_i^s}$ are calculated using 11 distinct Zeta functions where $s \in \{5, 4, 3, 2, 1, 0, -1, -2, -3, -4, -5\}$. These descriptors effectively capture the persistent spectral distribution of the electrostatic landscape. The resulting feature tensor for the E-Phys module is of size $30 \times 50 \times 11$, which serves as the input for a gradient boosting tree (GBT) architecture. By fusing these specialized geometric and electrostatic signatures, the GE-TopoLIA ensemble achieves a more comprehensive representation of the protein--ligand interaction.

 \section*{Acknowledgments}
 This work was supported in part by NIH grants R01AI164266 and R35GM148196, the University of Georgia,   the Georgia Research Alliance and  the Defense Advanced Research Projects Agency (DARPA) under Agreement No. HR00112969E087.
 MH was supported in part by the National Science Foundation (NSF grant DMS-2134231).
The views and conclusions contained in this document are those of the authors and should not be interpreted as representing the official policies, either expressed or implied, of the U.S. Government.

% \input{si_copy}
% \bibliographystyle{unsrtnat}
% \bibliography{reference}
% \end{document}
\bibliographystyle{unsrtnat}
\bibliography{reference}
\newpage

\begin{center}
    \Large \bfseries Supporting Information
\end{center}

\setcounter{section}{0}
\section{Summary of Datasets}
To rigorously evaluate the predictive and generative capabilities of the proposed TopoLIA model, we utilized benchmarks sourced from the PDBbind database \cite{liu2015pdb}. PDBbind serves as a curated, gold-standard resource that systematically collects experimentally determined 3D structures of protein--ligand complexes alongside their corresponding binding affinity data, providing the essential structural foundation for our pre-training and downstream scoring and docking tasks.

The dataset utilized for the self-supervised pre-training phase is a comprehensive compilation of protein--ligand complexes sourced from various versions of the PDBbind database, including PDBbind v.2007, v.2013, v.2016, and v.2020 \cite{liu2015pdb}. To ensure the integrity of the pre-training process and eliminate data leakage, a rigorous curation process was conducted to remove redundancies and labels. This resulted in a total of 19,462 non-overlapping complexes dedicated to pre-training.

For the standard scoring task, we evaluated TopoLIA on three widely recognized benchmarks: CASF-2007, CASF-2013, and CASF-2016 \cite{li2014comparative, cheng2009comparative, su2018comparative}. To maintain a strict evaluation protocol, the training set for each benchmark comprised the PDBbind refined set, with all entries belonging to the respective core sets explicitly excluded. The performance was then benchmarked on the independent core sets of CASF-2007, CASF-2013, and CASF-2016.  

The docking power evaluation was conducted using the CASF-2013 benchmark. This test set consists of 195 ligands, each accompanied by 100 decoy poses generated by three docking programs, GOLD v5.1, Surflex-Dock in SYBYL v8.1, and MOE v2011 \cite{li2014comparative, cheng2009comparative}. To prepare the training set for this task, 1,000 training poses were generated for each target ligand--receptor pair using GOLD v.5.6.33 \cite{jones1997development}. The pose structures and their corresponding GOLD scores are available for reproduction at \url{https://weilab.math.msu.edu/AGL-Score}. Following the methodology established in \cite{nguyen2019agl}, we conducted an individual training and testing process for each of the 195 test ligands in the CASF-2013 core set. The final success rate was computed as the percentage of the 195 ligands for which the model correctly identified a native pose as the top-ranked candidate.

To evaluate the model's generalization capabilities across targets with low sequence homology, we utilized the LBA30 and LBA60 datasets derived from the PDBbind v.2019 refined set. This set consists of 4,852 complexes, which are carefully curated based on several quality criteria, including a resolution of $\le 2.5$~\AA{}, an R-factor $\le 0.25$, and the absence of steric clashes or covalent bonds. Complexes with unreliable ligand binding data are excluded. To avoid overfitting, a protein sequence-based splitting strategy was implemented. Following the Atom3D \cite{townshend2020atom3d} partitioning method, the dataset with a 30\% sequence identity threshold (LBA30) contains 3,507 proteins in the training set, 466 proteins in the validation set, and 490 proteins in the test set. Similarly, the 60\% sequence identity threshold (LBA60) partition consists of 3,678 proteins for training, 460 for validation, and 460 for testing. This strategy prevents homologous proteins from appearing in both the training and test sets.

Finally, for the 10-fold cross-validation, binding affinity data for training and testing was retrieved from the PDBbind v.2020 database \cite{liu2015pdb}. The refined set is made up of 5,320 protein--ligand complexes with known binding affinities. The refined set provides a comprehensive and structurally diverse environment for robust model evaluation. Further details regarding all dataset splits are summarized in Table~\ref{Tab:summary}.

% contains binding affinity values expressed as either $K_d$ dissociation constants or $K_i$ inhibition constants, 
\begin{table}[ht]
\centering
\renewcommand{\arraystretch}{1.3} 
\begin{tabular}{p{3.5cm} p{4.5cm} l p{3.5cm}}
\toprule

\rowcolor{headerbeige} 
 & \textbf{Datasets} & \textbf{Training set} & \textbf{Test set} \\ 
\midrule

Pretraining \newline (self-supervised) & 
Combined PDBbind \newline (v.2020, 2007, 2013, 2016) \cite{liu2015pdb} & 
19,462  & 
/ \\ 
\midrule

\multirow{3}{3.5cm}{Scoring/Docking \newline (supervised fine-tuning)} 
 & CASF-2007 \cite{cheng2009comparative} & 1,105 & 195 \\ \cmidrule{2-4}
 & CASF-2013 \cite{li2014comparative} & 2,764 & 195  \\ \cmidrule{2-4}
 & CASF-2016 \cite{su2018comparative} & 3,772 & 285  \\ 
\midrule

\multirow{2}{3.5cm}{Generalization Splits \newline (Sequence Identity)} 
 & LBA30 (PDBbind v.2019) \cite{townshend2020atom3d} & 3,507 & 490 \\ \cmidrule{2-4}
 & LBA60 (PDBbind v.2019) \cite{townshend2020atom3d} & 3,678 & 460 \\ 
 \midrule

\multirow{1}{3.5cm}{Cross-Validation} 
 & PDBbind v.2020 refined \cite{liu2015pdb} & \multicolumn{2}{c}{5,320 (10-fold splits)} \\ 

\bottomrule
\end{tabular}
\caption{Summary of datasets used for pretraining, fine-tuning, and robust evaluations.}
\label{Tab:summary}
\end{table}

\section{Evaluation Metrics}

\subsection{Evaluation of Scoring Power}

To rigorously assess the scoring power of the TopoLIA model, we utilize the Pearson Correlation Coefficient (PCC) as the primary metric for linear dependency. The PCC measures the correlation between predicted and experimentally determined binding affinities and is defined as:

\begin{equation}
{ \rm PCC} = \frac{\sum_{i=1}^{n} (x_i - \bar{x})(y_i - \bar{y})}{\sqrt{\sum_{i=1}^{n} (x_i - \bar{x})^2 \sum_{i=1}^{n} (y_i - \bar{y})^2}}
\end{equation}

In this expression, $x_i$ represents the experimental value for the $i$-th sample, $\bar{x}$ denotes the mean of the experimental values, $y_i$ is the predicted value for the $i$-th sample, and $\bar{y}$ is the mean of the predicted values.

Furthermore, the Root Mean Squared Error (RMSE) is employed to quantify the absolute magnitude of the prediction error:

\begin{equation}
{\rm RMSE} = \sqrt{\frac{1}{n} \sum_{i=1}^{n} (y_i - \hat{y}_i)^2}
\end{equation}
where $y_i$ and $\hat{y}_i$ correspond to the predicted and actual binding affinities of the $i$-th sample, respectively.

\subsection{Evaluation of Docking Power}

The docking power evaluation measures a scoring function’s proficiency in distinguishing the native pose from an array of decoys generated by docking software. Within the benchmark parameters, a pose is deemed native if its root-mean-square deviation (RMSD) relative to the experimentally determined binding pose is less than 2 \AA.

Our validation is anchored to the CASF-2013 dataset, following the training and test sets delineated in the literature \cite{cang2018representability, meng2021persistent, nguyen2019agl}. The CASF-2013 benchmark provides 100 poses per ligand generated by three docking applications: GOLD v5.1, Surflex-Dock (SYBYL v8.1), and MOE v2011. The curated poses are accessible at \url{https://weilab.math.msu.edu/AGL-Score/}. 

Owing to molecular symmetries, a ligand may possess multiple native poses within the dataset; success is recorded if the model identifies any of these as the top-ranked candidate. The final docking power is calculated as the percentage of successful identifications:

\begin{equation}
\text{Docking Power} = \frac{\text{Number of successful complexes}}{\text{Total number of complexes in benchmark}} \times 100\%
\end{equation}

To assess structural similarity between the predicted docking mode ($A$) and the reference experimental structure ($B$), we utilize the Root Mean Square Deviation (RMSD):

\begin{equation}
{\rm RMSD} = \sqrt{\frac{1}{n} \sum_{i=1}^{n} \| A_i - B_i \|^2}
\end{equation}

where $A_i$ and $B_i$ represent the coordinates of the $i$-th atom in the docked and experimental structures, respectively, and $n$ denotes the total number of atoms compared. Lower RMSD values indicate higher predictive accuracy in reproducing known binding modes within the protein active site.

\section{Hyperparameter Selection and Optimization}
In this work, the Mean Squared Error (MSE) is applied as the loss function during both the pre-training and fine-tuning stages. The MSE is a standard metric used to quantify the variance between predicted and actual values by measuring the average of the squared differences:

\begin{equation}
{\rm MSE} = \frac{1}{n} \sum_{i=1}^{n} (y_i - \hat{y}_i)^2
\end{equation}
where $y_i$ is the actual value, $\hat{y}_i$ is the predicted value for the $i$-th data point, and $n$ represents the total number of samples. This objective function ensures that the model penalizes larger deviations more significantly, thereby refining the accuracy of the topological embeddings.

The TopoLIA framework was optimized using a two-stage training strategy: self-supervised pre-training followed by task-specific fine-tuning. The selection of hyperparameters was guided by the need to balance the expressive capacity of the latent interaction transformer with the stability required for convergence on complex biomolecular data. A comprehensive summary of these settings is provided in Table~\ref{tab:hyperparameters}.

During the pre-training stage, the model utilized an encoder-decoder architecture designed for a masked reconstruction task \cite{he2022masked}. The encoder consisted of 5 layers with a hidden dimension of 768 and 8 attention heads, while the decoder was comprised of 3 layers with a hidden dimension of 512 and 8 attention heads. To manage the high dimensionality of the topological sequence, the stalk dimensions were set to 96 for the encoder and 64 for the decoder. A masking ratio of 0.5 using a span-based approach was implemented to force the model to learn robust structural invariants from unlabeled protein--ligand interfaces.

For the fine-tuning stage, the decoder was discarded, and the pretrained encoder was integrated with a task-specific regression head for binding affinity prediction. This transition involved a reduction in the learning rate from $1 \times 10^{-4}$ to $5 \times 10^{-5}$ and a decrease in batch size from 64 to 32 to ensure stable gradients during supervised adaptation. Both stages utilized a low-rank optimization parameter ($r=8$) to enhance computational efficiency and prevent over-fitting.

\begin{table}[htbp]
\centering
\caption{Optimized Hyperparameter Settings for the TopoLIA Architecture.}
\label{tab:hyperparameters}
\begin{tabular}{lcc}
\toprule
Hyperparameter & Pre-training Stage & Fine-tuning Stage \\
\midrule
Learning Rate & $1 \times 10^{-4}$ & $5 \times 10^{-5}$ \\
Total Epochs & $100$ & $30$ \\
Batch Size & $64$ & $32$ \\
Encoder Hidden Dimension & $768$ & $768$ \\
Encoder Attention Heads & $8$ & $8$ \\
Encoder Layers & $5$ & $5$ \\
Encoder Stalk Dimension & $96$ & $96$ \\
Decoder Hidden Dimension & $512$ & N/A\\
Decoder Attention Heads & $8$ & N/A \\
Decoder Layers & $3$ & N/A \\
Decoder Stalk Dimension & $64$ & N/A \\
Masking Ratio / Type & $0.5$ / Span & N/A \\
Low-Rank Optimization ($r$) & $8$ & $8$ \\
Dropout & $0.1$ & $0.1$ \\
Weight Decay & $0.05$ & $0.05$ \\
Warm-up Rate & $0.05$ & $0.05$ \\
\bottomrule
\end{tabular}
\end{table}

The optimization process employed the AdamW optimizer with a weight decay of 0.05. To mitigate the risk of gradient instability, common in Transformer-based architectures, a linear warm-up strategy was implemented with a rate of 0.05. This ensured that the learning rate gradually increased over the first 5\% of the total training steps (100 epochs for pre-training and 30 epochs for fine-tuning) before following a standard decay schedule. Regularization was further maintained through a dropout rate of 0.1 across all layers.

To further enhance the predictive fidelity of the framework, the integrated TopoLIA model incorporates an E-Phys component designed to capture auxiliary electrostatic and physical descriptors. This component utilizes a gradient boosting tree (GBT) framework to model high-dimensional feature interactions that complement the primary geometric representations learned by the latent interaction transformer. The GBT model is configured with 4,000 estimators and a maximum depth of 7, allowing the architecture to capture intricate, non-linear relationships within the physical feature space. To ensure robust generalization and mitigate the risk of over-fitting, a learning rate of 0.01 is employed in conjunction with a subsampling rate of 0.7. Additionally, the model utilizes a square-root feature selection strategy at each split to increase the diversity of the individual trees within the ensemble.

\newpage

\section{Ablation Study}
\subsection{Ablation Study on Geometric and Electrostatic Modules}
The robust predictive scoring capabilities of our framework are driven by GE-TopoLIA, a heterogeneous ensemble that synergistically integrates two specialized modules: G-TopoLIA extracting geometric signatures via the latent interaction transformer, and the E-Phys module capturing electrostatic energy landscapes via gradient boosting trees. As quantitatively detailed in Table~\ref{tab:correlation_results}, this integrated GE-TopoLIA architecture consistently surpasses its standalone constituent modules across all evaluated benchmarks. This performance gain is clearly illustrated by the CASF-2007 dataset, where the ensemble approach achieved a leading PCC of 0.832 and an RMSE of 1.842 kcal/mol. In comparison, the standalone G-TopoLIA module yielded a PCC of 0.817 and an RMSE of 1.908 kcal/mol, while the E-Phys module produced a PCC of 0.805 and an RMSE of 1.970 kcal/mol. Such measured improvements validate that the fusion of geometric and electrostatic topological features provides a more comprehensive and accurate representation of protein--ligand binding affinity than models relying on a single feature domain. For clarity, the full GE-TopoLIA ensemble is hereafter denoted simply as TopoLIA.

\begin{table}[htbp]
\centering
\caption{PCC and RMSE (in kcal/mol) of the TopoLIA models on the three benchmarks of CASF-2007, CASF-2013
and CASF-2016.}
\label{tab:correlation_results}
\begin{tabular}{@{}lcccccc@{}}
\toprule
 & \multicolumn{2}{c}{G-TopoLIA} & \multicolumn{2}{c}{E-Phys} & \multicolumn{2}{c}{GE-TopoLIA} \\
\cmidrule(lr){2-3} \cmidrule(lr){4-5} \cmidrule(lr){6-7}
Benchmark & PCC & RMSE & PCC & RMSE & PCC & RMSE \\ \midrule
CASF-2007 & 0.817 & 1.908 & 0.805 & 1.970 & 0.832 & 1.842 \\
CASF-2013 & 0.798 & 1.931 & 0.778 & 2.041 & 0.814 & 1.908 \\
CASF-2016 & 0.857 & 1.597 & 0.824 & 1.769 & 0.861 & 1.604 \\ 
Average& 0.824 & 1.812 & 0.802 & 1.927 & 0.836 & 1.785 \\ \bottomrule
\end{tabular}
\end{table}

\subsection{Error Analysis of Predictive Outliers} 
To delineate the limitations of our framework, we conducted a systematic investigation into the complexes characterized by the largest predictive discrepancies. We compiled comprehensive performance Tables \ref{tab:full_2007}, \ref{tab:full_2013}, and \ref{tab:full_2016} across the CASF-2007, CASF-2013, and CASF-2016 benchmarks, sorted by decreasing predictive error. For the structural analysis presented here, we selected the most poorly predicted PDB entries from each benchmark as representatives to identify the fundamental factors constraining the model’s current predictive accuracy. These factors are primarily categorized into the structural simplification of the binding environment and chemical novelty relative to the training distribution.

A primary driver of predictive failure is the inherent discrepancy between the native crystallographic environment and the refined protein--ligand representations used for model training. A comparative analysis of representative outliers (PDB IDs: 2X00, 1Y6Q, and 3MYG) reveals that original structural records often contain essential components, such as coordinated water molecules, metal ions, and co-factors, that stabilize the binding pocket. As explicitly illustrated in the native structures in Figure \ref{fig:outlier_analysis}, these complexes features extensive hydration shells and ionic environments (represented by blue and yellow spheres). In high-affinity complexes, these elements frequently mediate critical charge-transfer effects or form water-bridge interactions that are indispensable for the observed binding affinity. However, as shown by the corresponding refined representations in Figure \ref{fig:outlier_analysis}, the TopoLIA model fundamentally relies on a highly simplified, desolvated protein--ligand input. By removing these critical environmental determinants during data refinement, the model is deprived of the requisite physicochemical features to justify high experimental affinities, thereby resulting in substantial under-prediction.

Beyond structural omissions, chemical novelty remains a formidable obstacle. We quantitatively evaluated the Tanimoto similarity between outlier ligands and the training distribution, as reported in Tables \ref{tab:full_2007}, \ref{tab:full_2013}, and \ref{tab:full_2016}. This analysis reveals that outliers with the largest errors in the CASF-2013 and CASF-2016 benchmarks, such as PDB IDs 3MYG and 2X00, possess exceptionally low similarity scores of 0.205 and 0.193, respectively. This chemical divergence indicates that the model struggles to generalize to novel molecular scaffolds that are poorly represented in the training set, emphasizing the need for broader data diversity or more robust transfer learning strategies in future iterations.

\section{Supplementary Figures}
\begin{figure}[!htbp] 
\centering  
\includegraphics[width=0.8\linewidth]{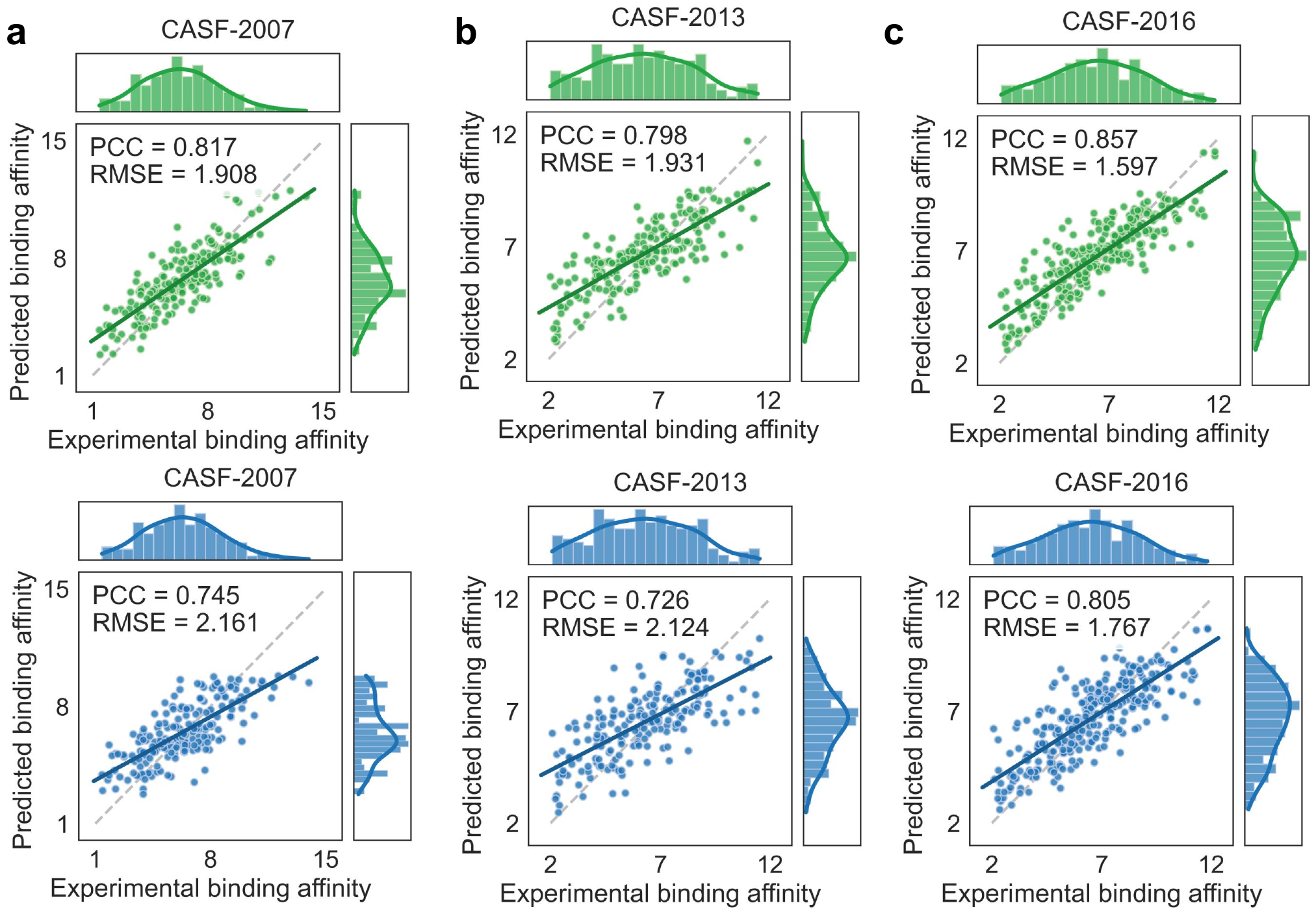} \caption{Scatter plots comparing experimental versus predicted binding affinities ($\text{p}K_d$) for the pretrained TopoLIA model (top) and a random initialization baseline (bottom) across three CASF benchmarks: (a) 2007, (b) 2013, and (c) 2016.}
\label{fig:compare} 
\end{figure}

\begin{figure}[!htbp] 
\centering \includegraphics[width=\linewidth]{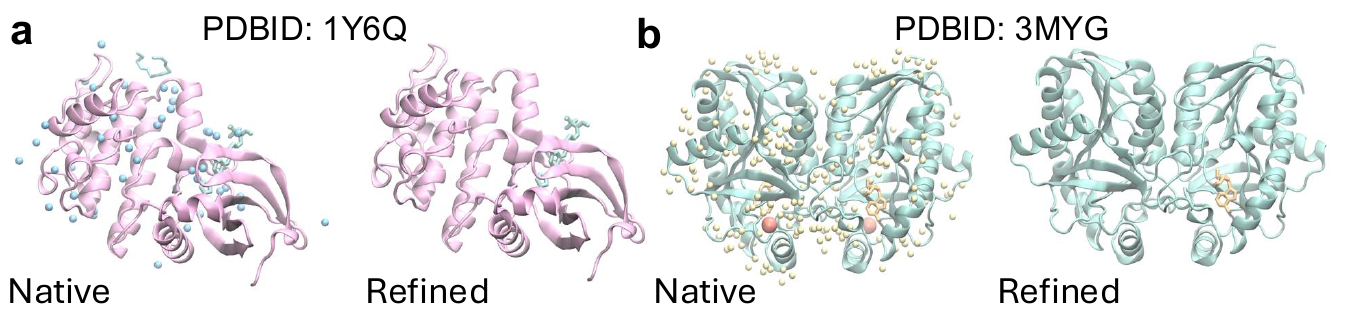} 
\caption{Comparative analysis of native and processed structural representations for outlier complexes. (a) Original crystal structures for PDB entries 1Y6Q. (b) Refined representations of PDB entries 1Y6Q utilized by the TopoLIA model. } 
\label{fig:outlier_analysis} 
\end{figure}

Figure~\ref{fig:PCC} details the scoring power of the TopoLIA model on the CASF-2007 and CASF-2013 benchmarks, featuring comparative correlation analyses against established state-of-the-art baselines alongside predictive scatter plots. To further validate our architectural design, Figure~\ref{fig:compare} provides a visual ablation study on the efficacy of self-supervised pretraining. By contrasting the prediction distributions of the pretrained model against a randomly initialized baseline, this analysis highlights the critical role of pretraining in achieving tighter predictive convergence. Finally, Figure~\ref{fig:outlier_analysis} illustrates a structural error analysis for predictive outliers (exemplified by PDB entry 1Y6Q). By visually comparing the native binding environments with their processed representations, this figure elucidates the structural omissions and environmental constraints underlying the observed predictive variance.

\begin{figure}[!htbp] 
\centering  \includegraphics[width=\linewidth]{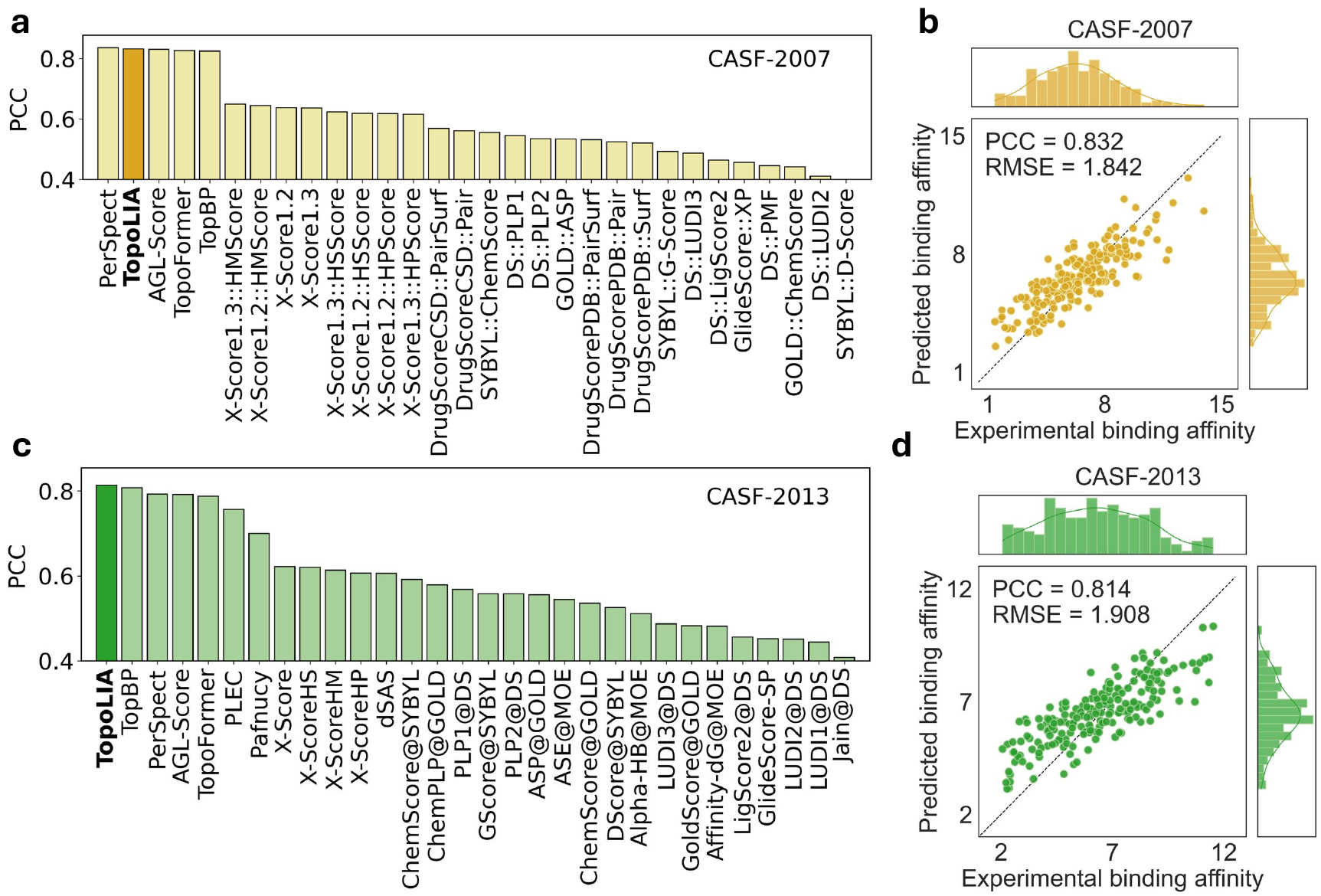} \caption{Comprehensive performance evaluation of the TopoLIA model for protein--ligand binding affinity scoring.
    (a, c) show a comparative analysis of the PCC between the proposed TopoLIA model and established cutting-edge scoring functions \cite{stepniewska2018development, cang2018representability, meng2021persistent, li2014comparative, su2018comparative, cheng2009comparative, wang2017improving, nguyen2019agl, wojcikowski2019development} for the CASF-2007, and CASF-2013 core sets, respectively. 
    (b, d) provide scatter plots illustrating the correlation between experimental and predicted binding affinities (in $pK_d$) for the same respective benchmarks. Each scatter panel includes the calculated PCC and RMSE (in kcal/mol).} \label{fig:PCC} 
% TopoLIA consistently achieves leading performance across all benchmarks, demonstrating its robustness against established competitors.

\end{figure}

\section{Supplementary Tables}

% Analysis of these detailed results reveals that the TopoLIA model maintains high predictive fidelity across a wide range of binding affinities, from weak ($pK_d < 3$) to very strong ($pK_d > 10$) interactions. By combining the geometric transport of the Copresheaf Transformer with the auxiliary physical descriptors provided by the E-Phys Gradient Boosting module, the framework effectively minimizes large-scale outliers. Even in cases where the Max Similarity to the training set is extremely low (e.g., 1SL3 or 2HDQ), the model demonstrates robust generalization, suggesting that the learned spectral-topological signatures capture fundamental principles of biomolecular recognition rather than relying on simple pattern matching.

Table \ref{tab:casf_comparison} presents a comprehensive quantitative evaluation of the proposed TopoLIA framework alongside other state-of-the-art \cite{stepniewska2018development, cang2018representability, meng2021persistent, li2014comparative, su2018comparative, cheng2009comparative, wang2017improving, nguyen2019agl, wojcikowski2019development} across the CASF-2007, CASF-2013, and CASF-2016 benchmarks for protein-ligand binding affinity predictions. The results demonstrate that TopoLIA consistently rank among the top-performing models, achieving superior PCC compared to traditional empirical scoring methods.

% Table \ref{tab:casf_comparison} presents a comprehensive quantitative comparison of the proposed TopoLIA framework with existing state-of-the-art scoring functions and learning-based methods \cite{stepniewska2018development, cang2018representability, meng2021persistent, li2014comparative, su2018comparative, cheng2009comparative, wang2017improving, nguyen2019agl, wojcikowski2019development} across the CASF-2007, CASF-2013, and CASF-2016 benchmarks for protein--ligand binding affinity prediction. The results show that TopoLIA consistently ranks among the top-performing models, achieving competitive or superior PCC values compared with traditional empirical scoring functions and recent topological learning approaches.

Table \ref{tab:docking_power} reports the docking-power evaluation on the CASF-2013 benchmark. This comparison assesses whether each scoring function can correctly identify near-native ligand poses from decoy conformations. TopoLIA achieves a high success rate for selecting the best pose, indicating that the proposed representation is not only effective for binding affinity prediction but also informative for pose-level ranking.

Table \ref{tab:lba_results} evaluates the generalization ability of TopoLIA under the Atom3D-curated LBA30 and LBA60 protein sequence identity splits. These benchmarks are designed to test model robustness when training and test proteins are separated by different sequence-similarity thresholds. The results demonstrate that TopoLIA achieves the best PCC and RMSE on both LBA30 and LBA60, supporting its ability to generalize beyond highly homologous protein--ligand complexes.

Tables \ref{tab:pretrain_scale}, \ref{tab:pst_vs_TopoLIA}, and \ref{tab:low_rank_ablation} further analyze the contribution of the major architectural and training components of TopoLIA. Table \ref{tab:pretrain_scale} examines the impact of self-supervised pretraining data scale on downstream CASF performance, showing that larger pretraining sets generally improve predictive accuracy. Table \ref{tab:pst_vs_TopoLIA} compares the baseline topological deep learning model with the proposed TopoLIA framework, demonstrating that incorporating the latent interaction attention consistently improves performance across the CASF benchmarks. Finally, Table \ref{tab:low_rank_ablation} studies the effect of the latent rank $r$ on both predictive performance and computational cost, identifying a practical bottleneck region that balances accuracy, training time, and GPU memory usage.

Table \ref{tab:full_2007}, Table \ref{tab:full_2013}, and Table \ref{tab:full_2016} provide the complete predictive profile for the TopoLIA model across the CASF-2007, CASF-2013, and CASF-2016 benchmarks. Each entry includes the PDB ID of the protein--ligand complex, the experimentally determined binding affinity ($pK_d$), and the corresponding value predicted by the integrated TopoLIA framework. To provide a clear overview of the model’s predictive limitations and strengths, the records in each table are sorted by descending absolute error. To facilitate a rigorous error analysis, we have also included the absolute error for each complex, alongside a max similarity metric that quantifies the structural relationship between the test ligand and the training set.

\begin{table*}[htbp]
\centering
\caption{The quantitative comparison of proposed models versus existing methods across the CASF-2007, CASF-2013, and CASF-2016 datasets. The results from other methods are taking from Refs\cite{stepniewska2018development, cang2018representability, meng2021persistent, li2014comparative, su2018comparative, cheng2009comparative, wang2017improving, nguyen2019agl, wojcikowski2019development}}.
\label{tab:casf_comparison}
\resizebox{\textwidth}{!}{
\begin{tabular}{r c r c r c}
\hline
\multicolumn{2}{c}{CASF-2007} & \multicolumn{2}{c}{CASF-2013} & \multicolumn{2}{c}{CASF-2016} \\
\hline
Methods & PCC & Methods & PCC & Methods & PCC \\
\hline
PerSpect & 0.836 & TopoLIA & 0.814 & TopoLIA & 0.861 \\ \hline
TopoLIA & 0.832 & TopBP & 0.808 & TopoFormer & 0.849 \\ \hline
AGL-Score & 0.830 & PerSpect & 0.793 & TopBP & 0.848 \\ \hline
TopoFormer & 0.826 & AGL-Score & 0.792 & PerSpect & 0.840 \\ \hline
TopBP & 0.825 & TopoFormer & 0.788 & AGL-Score & 0.835 \\ \hline
X-Score1.3::HMScore & 0.649 & PLEC & 0.757 & PLEC & 0.817 \\ \hline
X-Score1.2::HMScore & 0.644 & Pafnucy & 0.700 & DeltaVinaRF20 & 0.816 \\ \hline
X-Score1.2 & 0.637 & X-Score & 0.622 & Pafnucy & 0.780 \\ \hline
X-Score1.3 & 0.636 & X-ScoreHS & 0.620 & X-Score & 0.631 \\ \hline
X-Score1.3::HSScore & 0.624 & X-ScoreHM & 0.614 & X-ScoreHS & 0.629 \\ \hline
X-Score1.2::HSScore & 0.619 & X-ScoreHP & 0.607 & $\Delta$SAS & 0.625 \\ \hline
X-Score1.2::HPScore & 0.618 & dSAS & 0.606 & X-ScoreHP & 0.621 \\ \hline
X-Score1.3::HPScore & 0.616 & ChemScore@SYBYL & 0.592 & ASP@GOLD & 0.617 \\ \hline
DrugScoreCSD::PairSurf & 0.569 & ChemPLP@GOLD & 0.579 & ChemPLP@GOLD & 0.614 \\ \hline
DrugScoreCSD::Pair & 0.561 & PLP1@DS & 0.568 & X-ScoreHM & 0.609 \\ \hline
SYBYL::ChemScore & 0.555 & GScore@SYBYL & 0.558 & AutodockVina & 0.604 \\ \hline
DS::PLP1 & 0.545 & PLP2@DS & 0.558 & DrugScore2018 & 0.602 \\ \hline
DS::PLP2 & 0.535 & ASP@GOLD & 0.556 & DrugScoreCSD & 0.596 \\ \hline
GOLD::ASP & 0.534 & ASE@MOE & 0.544 & ASE@MOE & 0.591 \\ \hline
DrugScorePDB::PairSurf & 0.531 & ChemScore@GOLD & 0.536 & ChemScore@SYBYL & 0.590 \\ \hline
DrugScorePDB::Pair & 0.524 & DScore@SYBYL & 0.526 & PLP1@DS & 0.581 \\ \hline
DrugScorePDB::Surf & 0.520 & Alpha-HB@MOE & 0.511 & ChemScore@GOLD & 0.574 \\ \hline
SYBYL::G-Score & 0.492 & LUDI3@DS & 0.487 & G-Score@SYBYL & 0.572 \\ \hline
DS::LUDI3 & 0.487 & GoldScore@GOLD & 0.483 & Alpha-HB@MOE & 0.569 \\ \hline
DS::LigScore2 & 0.464 & Affinity-dG@MOE & 0.482 & PLP2@DS & 0.563 \\ \hline
GlideScore::XP & 0.457 & LigScore2@DS & 0.456 & Affinity-dG@MOE & 0.552 \\ \hline
DS::PMF & 0.445 & GlideScore-SP & 0.452 & LigScore2@DS & 0.540 \\ \hline
GOLD::ChemScore & 0.441 & LUDI2@DS & 0.451 & D-Score@SYBYL & 0.531 \\ \hline
DS::LUDI2 & 0.410 & LUDI1@DS & 0.444 & LUDI2@DS & 0.526 \\ \hline
SYBYL::D-Score & 0.392 & Jain@DS & 0.408 & GlideScore-SP & 0.513 \\ \hline
 &  &  &  & LUDI3@DS & 0.502 \\ \hline
\end{tabular}
}
\end{table*}

\begin{table}[htbp]
    \centering
    \caption{Comparison of docking power success rates on the CASF-2013 benchmark. The success rate was computed as the percentage of the 195 ligands for which the model correctly identified a native pose as the top-ranked candidate. The benchmark results are taken from the publications\cite{cang2017topologynet,huang2006iterative,wang2017improving,cheng2009comparative,li2013id,li2015improving,nguyen2019dg}.}
    \label{tab:docking_power}
    \begin{tabular}{lc}
        \toprule
        \textbf{Scoring Function} & \textbf{Success Rate (\%) of best pose} \\
        \midrule
        % Topoformer & 91 \\
        % AGL-Score & 90 \\
        TopoLIA & 88 \\
        $\Delta_{\text{vina}}\text{RF}_{20}$ & 87 \\
        AutoDock Vina & 85 \\
        ChemPLP@GOLD & 81 \\
        GlideScore-SP & 78 \\
        ChemScore@GOLD & 78 \\
        LigScore2@DS & 77 \\
        PLP1@DS & 77 \\
        Alpha-HB@MOE & 75 \\
        PLP2@DS & 74 \\
        GlideScore-XP & 74 \\
        ASP@GOLD & 72 \\
        GoldScore@GOLD & 71 \\
            LUDI2@DS & 66 \\
             LigScore1@DS & 65 \\
        Affinity-dG@MOE & 63 \\
        X-ScoreHM & 61 \\
         London-dG@MOE & 60 \\
             X-score & 60 \\
        ChemScore@SYBYL & 60 \\
        LUDI1@DS & 59 \\
         X-ScoreHS & 54 \\
           X-ScoreHP & 54 \\
        PMF@SYBYL & 52 \\
        PMF04@DS & 52 \\
        ASE@MOE & 51 \\
         LUDI3@DS & 49 \\
        Jain@DS & 48 \\
        G-Score@SYBYL & 45 \\
        PMF@DS & 45 \\
        D-Score@SYBYL & 18 \\
        \bottomrule
    \end{tabular}
\end{table}

\begin{table}[htbp]
\centering
\caption{Performance comparison of various methods on LBA datasets under different protein sequence identity split settings (LBA30 and LBA60). The RMSE values are reported in $pK_d$ units. The best results are highlighted in bold.}
\label{tab:lba_results}
\begin{tabular}{@{}lcccc@{}}
\toprule
\multirow{2}{*}{Method} & \multicolumn{2}{c}{LBA30} & \multicolumn{2}{c}{LBA60} \\
\cmidrule(lr){2-3} \cmidrule(lr){4-5}
 & PCC & RMSE & PCC & RMSE \\
\midrule
DeepDTA \cite{ozturk2018deepdta} & 0.472 & 1.866 & 0.666 & 1.762 \\
TAPE \cite{rao2019evaluating} & 0.338 & 1.890 & 0.568 & 1.633 \\
ProtTrans \cite{elnaggar2021prottrans} & 0.438 & 1.544 & 0.595 & 1.641 \\
Atom3D-CNN \cite{townshend2020atom3d} & 0.550 & 1.416 & 0.608 & 1.621 \\
Atom3D-ENN \cite{townshend2020atom3d} & 0.389 & 1.568 & 0.623 & 1.620 \\
Atom3D-GNN \cite{townshend2020atom3d} & 0.545 & 1.601 & 0.743 & 1.408 \\
Holoprot \cite{somnath2021multi} & 0.509 & 1.464 & 0.749 & 1.365 \\
ProNet \cite{wang2022learning} & 0.551 & 1.463 & 0.765 & 1.343 \\
DeepAffinity \cite{karimi2019deepaffinity} & 0.415 & 1.893 & -- & -- \\
SMT-DTA \cite{pei2022smt} & 0.458 & 1.574 & 0.758 & 1.347 \\
GeoSSL \cite{liu2022molecular} & 0.577 & 1.451 & -- & -- \\
Uni-Mol \cite{zhou2023uni} & 0.565 & 1.434 & 0.753 & 1.357 \\
BindNet \cite{feng2023protein} & 0.632 & 1.340 & 0.793 & 1.230 \\
SableBind \cite{li2025harnessing}& 0.579 & 1.527 & 0.802 & 1.246 \\
\textbf{TopoLIA} & \textbf{0.641} & \textbf{1.307} & \textbf{0.806} & \textbf{1.220} \\
\bottomrule
\end{tabular}
\end{table}

\begin{table}[htpb]
\centering
\caption{Impact of pretraining data scale on downstream predictive performance across three CASF benchmarks. The best results for each metric are highlighted in bold.}
\label{tab:pretrain_scale}
\begin{tabular}{lcccccc}
\toprule
\multirow{2}{*}{Pretraining Size} & \multicolumn{2}{c}{CASF-2007} & \multicolumn{2}{c}{CASF-2013} & \multicolumn{2}{c}{CASF-2016} \\
\cmidrule(lr){2-3} \cmidrule(lr){4-5} \cmidrule(lr){6-7}
& PCC  & RMSE  & PCC & RMSE  & PCC  & RMSE \\
\midrule
0 (Rand)      & 0.745 & 2.161 & 0.726 & 2.124 & 0.805 & 1.767 \\
1k            & 0.800 & 1.963 & 0.726 & 2.124 & 0.801 & 1.812 \\
5k            & 0.811 & 1.922 & 0.769 & 1.999 & 0.830 & 1.690 \\
10k           & 0.812 & 1.933 & 0.786 & 1.954 & 0.835 & 1.684 \\
19.5k (Full)  & \textbf{0.817} & \textbf{1.908} & \textbf{0.798} & \textbf{1.931} & \textbf{0.857} & \textbf{1.597} \\
\bottomrule
\end{tabular}
\end{table}

\begin{table}[htpb]
\centering
\caption{Performance comparison between the baseline topological deep learning (TopoSA) and the proposed topology-inspired latent interaction framework (TopoLIA). The inclusion of the latent interaction attention yields consistent improvements across all benchmarks.}
\label{tab:pst_vs_TopoLIA}
\begin{tabular}{lcccccc}
\toprule
\multirow{2}{*}{Model} & \multicolumn{2}{c}{CASF-2007} & \multicolumn{2}{c}{CASF-2013} & \multicolumn{2}{c}{CASF-2016} \\
\cmidrule(lr){2-3} \cmidrule(lr){4-5} \cmidrule(lr){6-7}
& PCC  & RMSE  & PCC & RMSE  & PCC& RMSE  \\
\midrule
TopoSA (Baseline) & 0.754 & 2.153 & 0.724 & 2.144 & 0.819 & 1.712 \\
TopoLIA (Ours)    & \textbf{0.817} & \textbf{1.908} & \textbf{0.798} & \textbf{1.931} & \textbf{0.857} & \textbf{1.597} \\
\bottomrule
\end{tabular}
\end{table}

\begin{table}[htpb]
\centering
\caption{Ablation study on the latent rank $r$, detailing predictive performance on CASF-2007 alongside computational costs (cumulative training time and peak GPU memory allocation). The optimal bottleneck region balancing accuracy and efficiency is highlighted in bold.}
\label{tab:low_rank_ablation}
\begin{tabular}{ccccc}
\toprule
Latent Rank ($r$) & PCC  & RMSE  & Training Time (Minutes) & Peak GPU Memory (GB) \\
\midrule
2  & 0.805 & 1.981 & 45.9  & 2.02  \\
4  & \textbf{0.808} & \textbf{1.959} & 69.9  & 2.53  \\
8  & \textbf{0.808} & 1.970 & 72.2  & 3.54  \\
16 & 0.783 & 2.062 & 74.0  & 5.56  \\
32 & 0.787 & 2.033 & 126.6 & 9.74  \\
64 & 0.765 & 2.133 & 184.8 & 18.08 \\
\bottomrule
\end{tabular}
\end{table}
\newpage

\begin{longtable}{lcccc}
\caption{Detailed scoring results for the CASF-2007 benchmark, sorted by descending absolute error.} \label{tab:full_2007} \\
\toprule
\textbf{PDB ID} & \textbf{Experimental ($pK_d$)} & \textbf{Predicted ($pK_d$)} & \textbf{Absolute Error} & \textbf{Max Similarity} \\
\midrule
\endfirsthead
\multicolumn{5}{c}{{\bfseries \tablename\ \thetable{} -- continued from previous page}} \\
\toprule
\textbf{PDB ID} & \textbf{Experimental ($pK_d$)} & \textbf{Predicted ($pK_d$)} & \textbf{Absolute Error} & \textbf{Max Similarity} \\
\midrule
\endhead
\midrule
\multicolumn{5}{r}{{Continued on next page}} \\
\endfoot
\bottomrule
\endlastfoot
1Y6Q & 11.70 & 7.625 & 4.075 & 0.508 \\
1SL3 & 11.85 & 8.228 & 3.622 & 0.000 \\
7CPA & 13.96 & 10.532 & 3.428 & 0.683 \\
1FLR & 10.00 & 6.641 & 3.359 & 0.323 \\
2DRC & 9.89 & 6.691 & 3.199 & 1.000 \\
1BMA & 4.59 & 7.696 & 3.106 & 0.486 \\
1Y1M & 1.82 & 4.678 & 2.858 & 0.889 \\
2FZC & 2.70 & 5.323 & 2.623 & 0.000 \\
1PB9 & 3.62 & 6.212 & 2.592 & 0.321 \\
2HDQ & 1.40 & 3.989 & 2.589 & 0.000 \\
1NC1 & 6.12 & 8.659 & 2.539 & 0.482 \\
1PPM & 5.80 & 8.309 & 2.509 & 0.662 \\
1F4E & 2.96 & 5.441 & 2.481 & 0.354 \\
2RKM & 3.90 & 6.358 & 2.458 & 0.935 \\
1BRA & 1.82 & 4.270 & 2.450 & 1.000 \\
1PXO & 8.70 & 6.269 & 2.431 & 0.435 \\
1U2Y & 1.74 & 4.130 & 2.390 & 1.000 \\
1MQ6 & 11.15 & 8.838 & 2.312 & 0.352 \\
1ZS0 & 6.15 & 8.438 & 2.288 & 0.000 \\
6RNT & 2.37 & 4.640 & 2.270 & 1.000 \\
1TYR & 7.00 & 4.805 & 2.195 & 0.000 \\
1LOQ & 3.70 & 5.883 & 2.183 & 0.837 \\
1PR5 & 3.92 & 6.071 & 2.151 & 0.646 \\
1BCU & 3.28 & 5.421 & 2.141 & 0.364 \\
1BXO & 10.00 & 7.896 & 2.104 & 0.358 \\
8CPA & 9.15 & 11.249 & 2.099 & 0.737 \\
1E66 & 9.89 & 7.816 & 2.074 & 0.241 \\
1NJE & 3.80 & 5.860 & 2.060 & 1.000 \\
1S39 & 7.70 & 5.653 & 2.047 & 0.625 \\
1A30 & 4.30 & 6.346 & 2.046 & 0.565 \\
1ZC9 & 3.22 & 5.266 & 2.046 & 0.000 \\
1NDW & 5.23 & 7.199 & 1.969 & 0.304 \\
2QWE & 7.48 & 5.643 & 1.837 & 0.692 \\
1SQA & 9.21 & 7.390 & 1.820 & 0.636 \\
1N2V & 4.08 & 5.894 & 1.814 & 0.236 \\
1SLG & 3.90 & 5.712 & 1.812 & 0.821 \\
1F4F & 4.62 & 6.410 & 1.790 & 0.361 \\
1JQ9 & 8.45 & 6.672 & 1.778 & 0.746 \\
2B1V & 5.74 & 7.493 & 1.753 & 0.463 \\
1GNI & 8.07 & 6.365 & 1.705 & 1.000 \\
2G94 & 9.52 & 7.865 & 1.655 & 0.333 \\
3PCE & 2.00 & 3.647 & 1.647 & 0.556 \\
5ER1 & 6.02 & 7.654 & 1.634 & 0.415 \\
1RE8 & 9.52 & 7.888 & 1.632 & 0.780 \\
1D7J & 3.30 & 4.866 & 1.566 & 0.286 \\
2G5U & 8.49 & 6.933 & 1.557 & 0.292 \\
3PCH & 5.40 & 3.857 & 1.543 & 1.000 \\
1NDY & 6.17 & 7.713 & 1.543 & 0.346 \\
1J17 & 5.22 & 6.759 & 1.539 & 1.000 \\
1XD1 & 7.92 & 6.404 & 1.516 & 0.719 \\
4TMN & 10.17 & 8.655 & 1.515 & 0.619 \\
1A69 & 5.30 & 6.813 & 1.513 & 1.000 \\
2AZR & 3.64 & 5.135 & 1.495 & 0.000 \\
1PZ5 & 5.40 & 6.860 & 1.460 & 0.483 \\
2CER & 9.22 & 7.763 & 1.457 & 0.660 \\
2J77 & 4.89 & 6.320 & 1.430 & 0.429 \\
1V48 & 7.80 & 6.379 & 1.421 & 0.400 \\
2CTC & 3.89 & 5.309 & 1.419 & 0.621 \\
4TIM & 2.16 & 3.577 & 1.417 & 0.419 \\
1P1Q & 4.89 & 6.284 & 1.394 & 0.440 \\
1AJP & 2.23 & 3.616 & 1.386 & 0.536 \\
4TLN & 3.72 & 5.095 & 1.375 & 0.538 \\
1Q7A & 7.19 & 5.830 & 1.360 & 0.286 \\
1JAQ & 4.48 & 5.833 & 1.353 & 0.431 \\
4ER2 & 9.30 & 7.951 & 1.349 & 1.000 \\
1FCX & 7.19 & 8.507 & 1.317 & 0.698 \\
1FTM & 7.61 & 6.316 & 1.294 & 0.440 \\
1Q8T & 4.76 & 6.043 & 1.283 & 0.268 \\
1IF7 & 10.52 & 9.251 & 1.269 & 1.000 \\
1E5A & 7.64 & 6.373 & 1.267 & 0.333 \\
1B7H & 8.02 & 6.796 & 1.224 & 0.895 \\
2QWB & 2.74 & 3.930 & 1.190 & 0.581 \\
1LOL & 6.39 & 5.207 & 1.183 & 0.000 \\
1B8O & 10.64 & 9.469 & 1.171 & 1.000 \\
2D0K & 5.02 & 6.166 & 1.146 & 0.694 \\
2BRB & 4.86 & 6.003 & 1.143 & 0.729 \\
1TOK & 2.47 & 3.571 & 1.101 & 0.333 \\
1UTP & 1.44 & 2.530 & 1.090 & 1.000 \\
1O3F & 7.96 & 6.873 & 1.087 & 1.000 \\
1VZQ & 7.44 & 8.524 & 1.084 & 0.514 \\
1GPK & 5.37 & 6.448 & 1.078 & 1.000 \\
2I0D & 12.10 & 11.026 & 1.074 & 0.864 \\
2C02 & 4.04 & 5.113 & 1.073 & 0.000 \\
2USN & 6.51 & 7.578 & 1.068 & 0.338 \\
1H23 & 8.35 & 9.412 & 1.062 & 1.000 \\
2AYR & 9.29 & 8.232 & 1.058 & 0.253 \\
2H3E & 5.70 & 6.751 & 1.051 & 0.000 \\
1X1Z & 11.06 & 10.009 & 1.051 & 0.511 \\
2FAI & 6.24 & 7.288 & 1.048 & 0.633 \\
2BOK & 6.55 & 7.582 & 1.032 & 0.623 \\
1L2S & 4.59 & 5.616 & 1.026 & 0.571 \\
1ELA & 6.36 & 7.360 & 1.000 & 0.534 \\
2B7D & 8.70 & 7.710 & 0.990 & 0.000 \\
1OLS & 5.82 & 4.842 & 0.978 & 1.000 \\
6STD & 8.64 & 9.611 & 0.971 & 0.490 \\
3PCJ & 7.22 & 6.266 & 0.954 & 0.516 \\
1ZOE & 7.40 & 6.449 & 0.951 & 0.500 \\
1U1B & 7.80 & 6.851 & 0.949 & 0.800 \\
1NHU & 5.66 & 6.576 & 0.916 & 0.310 \\
1FKB & 9.70 & 8.792 & 0.908 & 1.000 \\
2F80 & 8.18 & 9.081 & 0.901 & 1.000 \\
2STD & 9.85 & 10.743 & 0.893 & 1.000 \\
1U33 & 4.60 & 5.491 & 0.891 & 0.458 \\
1APW & 8.00 & 8.863 & 0.863 & 0.640 \\
1E1V & 4.92 & 5.782 & 0.862 & 1.000 \\
1TSY & 4.96 & 5.819 & 0.859 & 1.000 \\
2G8R & 3.99 & 4.833 & 0.843 & 0.000 \\
1NFY & 8.89 & 8.070 & 0.820 & 0.839 \\
3STD & 11.11 & 10.299 & 0.811 & 0.250 \\
1VFN & 5.60 & 4.797 & 0.803 & 0.519 \\
1TTM & 7.35 & 8.128 & 0.778 & 0.262 \\
1B9J & 5.96 & 6.733 & 0.773 & 0.789 \\
1KV1 & 5.94 & 6.700 & 0.760 & 0.283 \\
1ZVX & 9.22 & 8.470 & 0.750 & 0.000 \\
1XGJ & 6.00 & 5.282 & 0.718 & 0.500 \\
1DET & 4.30 & 4.997 & 0.697 & 0.000 \\
2FLB & 5.74 & 6.432 & 0.692 & 0.642 \\
1M0N & 2.22 & 2.908 & 0.688 & 0.583 \\
2BZ6 & 7.09 & 7.776 & 0.686 & 0.000 \\
1HI4 & 4.49 & 5.174 & 0.684 & 0.000 \\
2BAK & 7.43 & 8.093 & 0.663 & 0.354 \\
2BZZ & 6.43 & 5.767 & 0.663 & 0.000 \\
1TOI & 4.05 & 3.408 & 0.642 & 1.000 \\
2J78 & 6.42 & 5.790 & 0.630 & 1.000 \\
2CEQ & 7.28 & 6.650 & 0.630 & 1.000 \\
1J16 & 3.84 & 4.463 & 0.623 & 1.000 \\
1V2O & 4.73 & 4.118 & 0.612 & 1.000 \\
1IS0 & 7.00 & 6.408 & 0.592 & 0.648 \\
1M0Q & 3.89 & 3.309 & 0.581 & 0.625 \\
1G7Q & 6.06 & 6.635 & 0.575 & 0.457 \\
1DHI & 7.26 & 6.685 & 0.575 & 1.000 \\
2AOV & 7.04 & 6.472 & 0.568 & 0.309 \\
1B39 & 6.92 & 6.382 & 0.538 & 0.000 \\
1NNY & 7.66 & 7.140 & 0.520 & 0.753 \\
1FH9 & 6.43 & 5.914 & 0.516 & 0.452 \\
2F01 & 13.00 & 12.494 & 0.506 & 0.000 \\
2FDP & 7.59 & 8.087 & 0.497 & 0.321 \\
1JYS & 3.52 & 4.003 & 0.483 & 0.483 \\
2D3U & 6.92 & 7.380 & 0.460 & 0.000 \\
1HA2 & 5.54 & 6.000 & 0.460 & 1.000 \\
1A08 & 5.62 & 6.072 & 0.452 & 0.506 \\
1FZJ & 8.10 & 7.656 & 0.444 & 1.000 \\
1SYH & 6.31 & 5.868 & 0.442 & 1.000 \\
1FCZ & 9.22 & 8.779 & 0.441 & 0.667 \\
1O3P & 6.66 & 7.098 & 0.438 & 0.679 \\
1NJA & 6.31 & 5.888 & 0.422 & 1.000 \\
1AJQ & 4.31 & 3.895 & 0.415 & 0.400 \\
1FKN & 8.80 & 8.389 & 0.411 & 0.831 \\
1FH7 & 5.24 & 5.642 & 0.402 & 0.556 \\
1JQD & 5.16 & 5.556 & 0.396 & 1.000 \\
1SV3 & 4.74 & 4.345 & 0.395 & 0.000 \\
1O0H & 5.92 & 5.527 & 0.393 & 1.000 \\
2BAJ & 8.40 & 8.029 & 0.371 & 0.267 \\
1G7F & 5.47 & 5.830 & 0.360 & 0.506 \\
3GSS & 5.82 & 6.177 & 0.357 & 1.000 \\
1F5K & 3.74 & 4.096 & 0.356 & 1.000 \\
2ER9 & 7.40 & 7.742 & 0.342 & 0.483 \\
1OLU & 4.41 & 4.070 & 0.340 & 1.000 \\
1HFS & 8.70 & 8.983 & 0.283 & 0.358 \\
1ELB & 7.15 & 6.870 & 0.280 & 0.763 \\
1FH8 & 6.89 & 6.623 & 0.267 & 0.538 \\
2D1O & 7.70 & 7.947 & 0.247 & 1.000 \\
2BRM & 5.89 & 5.647 & 0.243 & 0.414 \\
2D3Z & 6.64 & 6.878 & 0.238 & 0.000 \\
1F4G & 6.48 & 6.716 & 0.236 & 0.341 \\
1NDZ & 8.11 & 8.345 & 0.235 & 0.286 \\
2QWD & 4.85 & 5.059 & 0.209 & 0.766 \\
1NVQ & 8.25 & 8.438 & 0.188 & 0.815 \\
1DF8 & 9.70 & 9.518 & 0.182 & 1.000 \\
1V16 & 3.87 & 4.051 & 0.181 & 1.000 \\
1FZK & 8.40 & 8.219 & 0.181 & 1.000 \\
1K9S & 6.52 & 6.689 & 0.169 & 0.660 \\
2CGR & 7.28 & 7.112 & 0.168 & 0.305 \\
2GSS & 4.94 & 4.782 & 0.158 & 0.378 \\
1A1B & 6.40 & 6.245 & 0.155 & 0.677 \\
1HK4 & 5.31 & 5.461 & 0.151 & 0.750 \\
1TMN & 7.30 & 7.448 & 0.148 & 0.486 \\
1RNT & 5.19 & 5.334 & 0.144 & 0.000 \\
1AI5 & 3.72 & 3.587 & 0.133 & 0.613 \\
1AVN & 3.90 & 4.015 & 0.115 & 0.367 \\
1PBQ & 6.27 & 6.374 & 0.104 & 0.333 \\
1FD0 & 8.40 & 8.320 & 0.080 & 0.756 \\
1YDT & 7.32 & 7.253 & 0.067 & 0.607 \\
2AOU & 7.73 & 7.795 & 0.065 & 0.413 \\
1OM1 & 6.77 & 6.821 & 0.051 & 0.364 \\
2CET & 8.02 & 7.974 & 0.046 & 0.660 \\
10GS & 6.40 & 6.366 & 0.034 & 0.457 \\
1KV5 & 4.22 & 4.254 & 0.034 & 1.000 \\
1TOJ & 3.39 & 3.362 & 0.028 & 1.000 \\
1UWT & 5.97 & 5.998 & 0.028 & 1.000 \\
1K4G & 5.85 & 5.826 & 0.024 & 1.000 \\
1M2Q & 6.10 & 6.084 & 0.016 & 0.444 \\
1TRD & 5.40 & 5.407 & 0.007 & 1.000 \\
1FKI & 7.00 & 6.994 & 0.006 & 0.231 \\
1D09 & 7.57 & 7.569 & 0.001 & 1.000 \\
\end{longtable}

\begin{longtable}{lcccc}
\caption{Detailed scoring results for the CASF-2013 benchmark, sorted by descending absolute error.} \label{tab:full_2013} \\
\toprule
\textbf{PDB ID} & \textbf{Experimental ($pK_d$)} & \textbf{Predicted ($pK_d$)} & \textbf{Absolute Error} & \textbf{Max Similarity} \\
\midrule
\endfirsthead
\multicolumn{5}{c}{{\bfseries \tablename\ \thetable{} -- continued from previous page}} \\
\toprule
\textbf{PDB ID} & \textbf{Experimental ($pK_d$)} & \textbf{Predicted ($pK_d$)} & \textbf{Absolute Error} & \textbf{Max Similarity} \\
\midrule
\endhead
\midrule
\multicolumn{5}{r}{{Continued on next page}} \\
\endfoot
\bottomrule
\endlastfoot
3MYG & 10.70 & 6.988 & 3.712 & 0.206 \\
2X00 & 11.33 & 7.848 & 3.482 & 0.194 \\
3LKA & 2.82 & 6.118 & 3.298 & 0.480 \\
2XDL & 3.10 & 6.214 & 3.114 & 0.487 \\
1IGJ & 10.00 & 7.023 & 2.977 & 0.671 \\
2YMD & 3.16 & 6.071 & 2.911 & 1.000 \\
3AO4 & 2.07 & 4.898 & 2.828 & 0.619 \\
1PS3 & 2.28 & 5.102 & 2.822 & 1.000 \\
3MFV & 2.52 & 5.208 & 2.688 & 0.750 \\
3SU3 & 9.13 & 6.443 & 2.687 & 1.000 \\
3F3A & 4.19 & 6.753 & 2.563 & 1.000 \\
2PCP & 8.70 & 6.145 & 2.555 & 0.310 \\
1LOQ & 3.70 & 6.235 & 2.535 & 0.837 \\
4GID & 10.77 & 8.376 & 2.394 & 0.553 \\
1E66 & 9.89 & 7.517 & 2.373 & 0.266 \\
2J62 & 11.34 & 8.969 & 2.371 & 0.768 \\
4GQQ & 2.89 & 5.235 & 2.345 & 0.325 \\
1LBK & 3.18 & 5.475 & 2.295 & 1.000 \\
1OS0 & 6.03 & 8.284 & 2.254 & 1.000 \\
1MQ6 & 11.15 & 8.906 & 2.244 & 0.352 \\
3UTU & 10.92 & 8.716 & 2.204 & 0.477 \\
3B3S & 2.55 & 4.748 & 2.198 & 1.000 \\
3KGP & 2.57 & 4.765 & 2.195 & 0.520 \\
2CBJ & 8.27 & 6.087 & 2.183 & 1.000 \\
1N2V & 4.08 & 6.251 & 2.171 & 0.246 \\
3ACW & 4.76 & 6.904 & 2.144 & 0.214 \\
3NQ3 & 3.78 & 5.891 & 2.111 & 1.000 \\
1BCU & 3.28 & 5.374 & 2.094 & 1.000 \\
3OZT & 4.13 & 6.214 & 2.084 & 0.685 \\
2VO5 & 4.89 & 6.960 & 2.070 & 0.268 \\
2W66 & 4.05 & 6.044 & 1.994 & 0.686 \\
3URI & 9.00 & 7.060 & 1.940 & 0.456 \\
3IVG & 4.30 & 6.235 & 1.935 & 0.723 \\
2QMJ & 4.21 & 6.128 & 1.918 & 1.000 \\
1Q8T & 4.76 & 6.665 & 1.905 & 0.292 \\
1P1Q & 4.89 & 6.777 & 1.887 & 1.000 \\
2ZX6 & 10.60 & 8.714 & 1.886 & 0.525 \\
2VW5 & 8.52 & 6.645 & 1.875 & 0.429 \\
1JYQ & 8.70 & 6.893 & 1.807 & 0.411 \\
1F8C & 7.40 & 5.606 & 1.794 & 1.000 \\
3FCQ & 2.77 & 4.563 & 1.793 & 0.559 \\
3UDH & 2.85 & 4.612 & 1.762 & 0.231 \\
1SQA & 9.21 & 7.462 & 1.748 & 0.636 \\
3AG9 & 8.05 & 6.316 & 1.734 & 0.901 \\
3F80 & 4.22 & 5.945 & 1.725 & 0.615 \\
3FK1 & 2.62 & 4.324 & 1.704 & 1.000 \\
1XD0 & 7.12 & 5.422 & 1.698 & 0.604 \\
3KV2 & 7.32 & 5.628 & 1.692 & 0.528 \\
2WBG & 4.45 & 6.142 & 1.692 & 1.000 \\
2JDY & 4.37 & 6.041 & 1.671 & 1.000 \\
2ZCQ & 8.82 & 7.152 & 1.668 & 0.609 \\
2QBP & 8.40 & 6.763 & 1.637 & 0.750 \\
3KWA & 4.08 & 5.707 & 1.627 & 0.929 \\
3B3W & 4.19 & 5.796 & 1.606 & 1.000 \\
2G70 & 7.77 & 6.185 & 1.585 & 0.393 \\
1A30 & 4.30 & 5.881 & 1.581 & 0.650 \\
4DE2 & 4.12 & 5.701 & 1.581 & 0.660 \\
4TMN & 10.17 & 8.605 & 1.565 & 0.619 \\
2BRB & 4.86 & 6.416 & 1.556 & 0.729 \\
1QI0 & 2.35 & 3.892 & 1.542 & 1.000 \\
2V00 & 3.66 & 5.192 & 1.532 & 0.400 \\
1F8D & 3.40 & 4.881 & 1.481 & 0.822 \\
2ZJW & 7.70 & 6.246 & 1.454 & 0.317 \\
3MSS & 4.66 & 6.088 & 1.428 & 0.526 \\
1Z95 & 7.12 & 8.522 & 1.402 & 0.561 \\
2OBF & 8.85 & 7.464 & 1.386 & 0.509 \\
3IMC & 2.96 & 4.343 & 1.383 & 0.386 \\
2XY9 & 9.19 & 7.810 & 1.380 & 1.000 \\
3U9Q & 4.38 & 5.754 & 1.374 & 1.000 \\
3L7B & 2.40 & 3.763 & 1.363 & 0.625 \\
2PQ9 & 8.11 & 6.756 & 1.354 & 1.000 \\
3CFT & 4.19 & 5.498 & 1.308 & 0.439 \\
1GPK & 5.37 & 6.674 & 1.304 & 0.431 \\
3L3N & 8.18 & 6.879 & 1.301 & 1.000 \\
3PE2 & 9.76 & 8.483 & 1.277 & 0.547 \\
2Y5H & 5.79 & 7.066 & 1.276 & 0.736 \\
3CJ2 & 4.85 & 6.106 & 1.256 & 0.558 \\
1YC1 & 6.17 & 7.425 & 1.255 & 0.328 \\
2YKI & 9.46 & 8.266 & 1.194 & 0.272 \\
3G0W & 9.52 & 8.336 & 1.184 & 0.284 \\
3I3B & 2.23 & 3.413 & 1.183 & 1.000 \\
3F3E & 7.70 & 6.519 & 1.181 & 1.000 \\
4DJR & 11.52 & 10.341 & 1.179 & 0.667 \\
1ZEA & 5.22 & 6.393 & 1.173 & 0.589 \\
2X97 & 5.66 & 6.816 & 1.156 & 0.481 \\
2HB1 & 3.80 & 4.948 & 1.148 & 0.000 \\
1U1B & 7.80 & 6.653 & 1.147 & 0.800 \\
1F8B & 5.40 & 4.315 & 1.085 & 1.000 \\
3OWJ & 6.07 & 7.133 & 1.063 & 0.288 \\
3EBP & 5.91 & 4.852 & 1.058 & 0.297 \\
2QFT & 5.26 & 4.213 & 1.047 & 1.000 \\
3DXG & 2.40 & 3.444 & 1.044 & 0.837 \\
2P4Y & 9.00 & 7.957 & 1.043 & 0.337 \\
2XNB & 6.83 & 7.871 & 1.041 & 0.544 \\
3F17 & 8.63 & 7.598 & 1.032 & 0.513 \\
3VD4 & 4.82 & 3.790 & 1.030 & 1.000 \\
3N7A & 3.70 & 4.725 & 1.025 & 1.000 \\
1O3F & 7.96 & 6.981 & 0.979 & 1.000 \\
4G8M & 7.89 & 6.916 & 0.974 & 1.000 \\
2ZWZ & 7.79 & 6.825 & 0.965 & 0.385 \\
3SU5 & 5.58 & 6.542 & 0.962 & 1.000 \\
2YGE & 5.06 & 6.019 & 0.959 & 0.506 \\
3G2N & 4.09 & 5.041 & 0.951 & 0.564 \\
3L4U & 7.52 & 6.571 & 0.949 & 0.659 \\
3GE7 & 8.70 & 7.752 & 0.948 & 0.507 \\
4DEW & 7.00 & 6.071 & 0.929 & 0.778 \\
3CYX & 8.00 & 8.925 & 0.925 & 1.000 \\
3EHY & 5.85 & 6.766 & 0.916 & 0.558 \\
4DJV & 6.72 & 7.627 & 0.907 & 0.673 \\
3GBB & 6.90 & 7.804 & 0.904 & 0.652 \\
3VH9 & 6.24 & 5.337 & 0.903 & 0.417 \\
3GCS & 7.25 & 8.134 & 0.884 & 1.000 \\
2JDM & 5.40 & 6.278 & 0.878 & 1.000 \\
2V7A & 8.30 & 7.424 & 0.876 & 0.340 \\
3ZSX & 3.28 & 4.152 & 0.872 & 0.608 \\
1VSO & 4.72 & 5.588 & 0.868 & 0.345 \\
1UTO & 2.27 & 3.136 & 0.866 & 1.000 \\
3K5V & 6.30 & 7.136 & 0.836 & 0.322 \\
2R23 & 3.72 & 4.554 & 0.834 & 0.741 \\
3NOX & 8.66 & 7.828 & 0.832 & 0.380 \\
3UO4 & 6.52 & 7.348 & 0.828 & 0.679 \\
3G2Z & 2.36 & 3.173 & 0.813 & 0.231 \\
3UEU & 5.24 & 6.048 & 0.808 & 1.000 \\
1H23 & 8.35 & 9.155 & 0.805 & 1.000 \\
2ZXD & 5.22 & 6.023 & 0.803 & 0.480 \\
1W3K & 4.30 & 5.098 & 0.798 & 0.659 \\
3GNW & 9.10 & 8.316 & 0.784 & 0.635 \\
1LOR & 11.06 & 10.280 & 0.780 & 1.000 \\
1KEL & 7.28 & 6.505 & 0.775 & 0.500 \\
2VVN & 7.30 & 6.538 & 0.762 & 0.667 \\
1LOL & 6.39 & 5.641 & 0.749 & 0.000 \\
3N86 & 5.64 & 6.384 & 0.744 & 0.450 \\
3NW9 & 9.00 & 8.264 & 0.736 & 0.844 \\
3PXF & 4.43 & 5.146 & 0.716 & 1.000 \\
2QBR & 6.33 & 7.046 & 0.716 & 0.661 \\
2WTV & 8.74 & 8.027 & 0.713 & 0.361 \\
2X0Y & 4.60 & 5.312 & 0.712 & 0.422 \\
1U33 & 4.60 & 5.309 & 0.709 & 0.511 \\
2ZCR & 6.87 & 7.562 & 0.692 & 0.769 \\
3F3C & 6.02 & 6.699 & 0.679 & 0.724 \\
2XB8 & 7.59 & 6.914 & 0.676 & 1.000 \\
3SU2 & 7.35 & 6.681 & 0.669 & 1.000 \\
3OV1 & 5.20 & 5.839 & 0.639 & 0.931 \\
1W3L & 6.28 & 5.651 & 0.629 & 0.625 \\
2JDU & 6.72 & 6.111 & 0.609 & 1.000 \\
2XHM & 6.80 & 7.403 & 0.603 & 0.507 \\
3BKK & 6.08 & 6.673 & 0.593 & 0.741 \\
3B68 & 8.40 & 8.943 & 0.543 & 0.719 \\
3E93 & 8.85 & 8.309 & 0.541 & 0.461 \\
2J78 & 6.42 & 5.882 & 0.538 & 1.000 \\
1SLN & 6.64 & 7.160 & 0.520 & 0.440 \\
3ZSO & 5.12 & 5.619 & 0.499 & 0.824 \\
3JVS & 6.54 & 6.060 & 0.480 & 0.375 \\
2FVD & 8.52 & 8.049 & 0.471 & 0.277 \\
3FV1 & 9.30 & 8.829 & 0.471 & 1.000 \\
3HUC & 5.99 & 6.456 & 0.466 & 0.286 \\
4DES & 5.85 & 6.309 & 0.459 & 1.000 \\
3D4Z & 4.89 & 5.337 & 0.447 & 1.000 \\
10GS & 6.40 & 6.827 & 0.427 & 0.466 \\
3GY4 & 5.10 & 4.675 & 0.425 & 0.571 \\
3S8O & 6.85 & 6.445 & 0.405 & 1.000 \\
3UEX & 6.92 & 6.528 & 0.392 & 1.000 \\
1OYT & 7.24 & 7.588 & 0.348 & 0.623 \\
2XBV & 8.43 & 8.087 & 0.343 & 0.618 \\
3COY & 6.02 & 6.357 & 0.337 & 0.836 \\
3BFU & 6.27 & 6.603 & 0.333 & 1.000 \\
4DE1 & 5.96 & 5.628 & 0.332 & 0.459 \\
2IWX & 6.68 & 7.008 & 0.328 & 0.976 \\
1R5Y & 6.46 & 6.151 & 0.309 & 0.625 \\
2CET & 8.02 & 8.317 & 0.297 & 1.000 \\
2WCA & 5.60 & 5.891 & 0.291 & 1.000 \\
2VL4 & 6.01 & 5.724 & 0.286 & 0.481 \\
2YFE & 6.63 & 6.886 & 0.256 & 0.314 \\
1HNN & 6.24 & 6.470 & 0.230 & 0.606 \\
1W4O & 5.22 & 5.014 & 0.206 & 1.000 \\
2GSS & 4.94 & 5.143 & 0.203 & 0.378 \\
3PWW & 7.32 & 7.517 & 0.197 & 1.000 \\
3L4W & 6.00 & 5.811 & 0.189 & 0.667 \\
3BPC & 4.80 & 4.984 & 0.184 & 0.452 \\
3OE5 & 6.88 & 7.057 & 0.177 & 0.594 \\
3DD0 & 9.00 & 9.164 & 0.164 & 1.000 \\
2OLE & 7.25 & 7.093 & 0.157 & 0.263 \\
3EJR & 8.57 & 8.415 & 0.155 & 0.761 \\
1N1M & 5.70 & 5.558 & 0.142 & 0.296 \\
1O5B & 5.77 & 5.647 & 0.123 & 1.000 \\
2D3U & 6.92 & 6.797 & 0.123 & 0.000 \\
2VOT & 7.14 & 7.262 & 0.122 & 1.000 \\
3MUZ & 3.46 & 3.576 & 0.116 & 1.000 \\
2D1O & 7.70 & 7.596 & 0.104 & 1.000 \\
2X8Z & 7.96 & 7.908 & 0.052 & 1.000 \\
2XYS & 7.42 & 7.374 & 0.046 & 0.152 \\
2WEG & 6.50 & 6.514 & 0.014 & 0.538 \\
1HFS & 8.70 & 8.712 & 0.012 & 0.322 \\
1NVQ & 8.25 & 8.240 & 0.010 & 0.815 \\
1Q8U & 5.96 & 5.960 & 0.000 & 0.607 \\
\end{longtable}

\begin{longtable}{lcccc}
\caption{Detailed scoring results for the CASF-2016 benchmark, sorted by descending absolute error.} \label{tab:full_2016} \\
\toprule
\textbf{PDB ID} & \textbf{Experimental ($pK_d$)} & \textbf{Predicted ($pK_d$)} & \textbf{Absolute Error} & \textbf{Max Similarity} \\
\midrule
\endfirsthead
\multicolumn{5}{c}{{\bfseries \tablename\ \thetable{} -- continued from previous page}} \\
\toprule
\textbf{PDB ID} & \textbf{Experimental ($pK_d$)} & \textbf{Predicted ($pK_d$)} & \textbf{Absolute Error} & \textbf{Max Similarity} \\
\midrule
\endhead
\midrule
\multicolumn{5}{r}{{Continued on next page}} \\
\endfoot
\bottomrule
\endlastfoot
2X00 & 11.33 & 8.027 & 3.303 & 0.194 \\
3MYG & 10.70 & 7.455 & 3.245 & 0.206 \\
3OZT & 4.13 & 7.226 & 3.096 & 0.612 \\
5DWR & 11.22 & 8.187 & 3.033 & 0.274 \\
2BR1 & 5.14 & 8.100 & 2.960 & 0.414 \\
2BRB & 4.86 & 7.751 & 2.891 & 0.571 \\
2XDL & 3.10 & 5.982 & 2.882 & 0.487 \\
2YMD & 3.16 & 6.033 & 2.873 & 1.000 \\
1PS3 & 2.28 & 5.088 & 2.808 & 1.000 \\
3DX1 & 3.58 & 6.379 & 2.799 & 0.333 \\
5C2H & 11.09 & 8.354 & 2.736 & 0.446 \\
1E66 & 9.89 & 7.161 & 2.729 & 0.268 \\
3LKA & 2.82 & 5.426 & 2.606 & 0.607 \\
4U4S & 2.92 & 5.429 & 2.509 & 0.545 \\
4QD6 & 8.64 & 6.170 & 2.470 & 0.275 \\
3AO4 & 2.07 & 4.499 & 2.429 & 0.619 \\
4GID & 10.77 & 8.393 & 2.377 & 0.506 \\
1BCU & 3.28 & 5.548 & 2.268 & 1.000 \\
1MQ6 & 11.15 & 8.925 & 2.225 & 0.352 \\
4MGD & 4.69 & 6.895 & 2.205 & 0.367 \\
3ARU & 3.22 & 5.420 & 2.200 & 1.000 \\
3U8N & 10.17 & 8.054 & 2.116 & 0.423 \\
3UTU & 10.92 & 8.825 & 2.095 & 0.477 \\
3F3A & 4.19 & 6.253 & 2.063 & 1.000 \\
3BV9 & 5.36 & 7.422 & 2.062 & 0.421 \\
5A7B & 3.57 & 5.631 & 2.061 & 0.481 \\
3RR4 & 4.55 & 6.546 & 1.996 & 0.667 \\
2VW5 & 8.52 & 6.560 & 1.960 & 0.467 \\
1Q8T & 4.76 & 6.715 & 1.955 & 0.396 \\
1P1Q & 4.89 & 6.839 & 1.949 & 1.000 \\
2POG & 9.54 & 7.609 & 1.931 & 0.296 \\
1Y6R & 10.11 & 8.215 & 1.895 & 0.654 \\
4RFM & 10.05 & 8.164 & 1.886 & 0.205 \\
3IVG & 4.30 & 6.183 & 1.883 & 0.723 \\
4EKY & 3.52 & 5.372 & 1.852 & 0.692 \\
4JSZ & 2.30 & 4.117 & 1.817 & 1.000 \\
4EOR & 6.30 & 8.078 & 1.778 & 1.000 \\
3GV9 & 2.12 & 3.874 & 1.754 & 0.353 \\
2QBP & 8.40 & 6.647 & 1.753 & 0.750 \\
3ACW & 4.76 & 6.490 & 1.730 & 0.282 \\
1SQA & 9.21 & 7.496 & 1.714 & 0.534 \\
3F3E & 7.70 & 5.998 & 1.702 & 1.000 \\
3UDH & 2.85 & 4.522 & 1.672 & 0.231 \\
3UUO & 7.96 & 6.317 & 1.643 & 0.244 \\
4TMN & 10.17 & 8.528 & 1.642 & 0.551 \\
1A30 & 4.30 & 5.924 & 1.624 & 0.650 \\
2WN9 & 8.52 & 6.917 & 1.603 & 0.792 \\
2WBG & 4.45 & 6.031 & 1.581 & 1.000 \\
3OZS & 5.33 & 6.895 & 1.565 & 0.554 \\
3URI & 9.00 & 7.435 & 1.565 & 0.456 \\
3KGP & 2.57 & 4.098 & 1.528 & 0.520 \\
2V00 & 3.66 & 5.169 & 1.509 & 0.400 \\
2ZCQ & 8.82 & 7.322 & 1.498 & 0.609 \\
4PCS & 7.85 & 6.357 & 1.493 & 0.383 \\
4TY7 & 9.52 & 8.037 & 1.483 & 0.524 \\
4DE2 & 4.12 & 5.599 & 1.479 & 0.660 \\
3KWA & 4.08 & 5.557 & 1.477 & 0.929 \\
1S38 & 5.15 & 6.625 & 1.475 & 0.543 \\
4KZ6 & 3.10 & 4.566 & 1.466 & 0.391 \\
1QKT & 9.04 & 7.585 & 1.455 & 1.000 \\
2ZB1 & 6.32 & 7.768 & 1.448 & 0.507 \\
3L7B & 2.40 & 3.845 & 1.445 & 0.625 \\
4TWP & 10.00 & 8.556 & 1.444 & 1.000 \\
4JIA & 9.22 & 7.790 & 1.430 & 0.325 \\
2Y5H & 5.79 & 7.219 & 1.429 & 0.736 \\
4IH5 & 4.11 & 5.529 & 1.419 & 0.294 \\
3G0W & 9.52 & 8.103 & 1.417 & 0.284 \\
3UI7 & 9.00 & 7.595 & 1.405 & 0.230 \\
3MSS & 4.66 & 6.044 & 1.384 & 0.526 \\
4F3C & 11.82 & 10.437 & 1.383 & 1.000 \\
4AGN & 3.97 & 5.348 & 1.378 & 0.714 \\
5ABA & 2.98 & 4.314 & 1.334 & 0.322 \\
4W9L & 5.02 & 6.351 & 1.331 & 0.838 \\
2AL5 & 8.40 & 7.072 & 1.328 & 0.700 \\
1GPK & 5.37 & 6.678 & 1.308 & 0.197 \\
2XNB & 6.83 & 8.137 & 1.307 & 0.544 \\
3UO4 & 6.52 & 7.817 & 1.297 & 0.679 \\
3U9Q & 4.38 & 5.674 & 1.294 & 1.000 \\
1SYI & 5.44 & 6.723 & 1.283 & 1.000 \\
2XJ7 & 6.66 & 5.393 & 1.267 & 1.000 \\
2P15 & 10.30 & 9.037 & 1.263 & 0.455 \\
3B65 & 9.27 & 8.032 & 1.238 & 0.597 \\
3RSX & 4.41 & 5.627 & 1.217 & 0.297 \\
2W66 & 4.05 & 5.267 & 1.217 & 0.686 \\
1Z95 & 7.12 & 8.330 & 1.210 & 0.561 \\
2P4Y & 9.00 & 7.795 & 1.205 & 0.337 \\
1Z6E & 9.72 & 8.520 & 1.200 & 0.474 \\
4QAC & 9.40 & 8.228 & 1.172 & 0.280 \\
3B1M & 8.48 & 7.311 & 1.169 & 0.248 \\
3G31 & 2.89 & 4.048 & 1.158 & 0.281 \\
4M0Z & 5.19 & 6.332 & 1.142 & 0.815 \\
1U1B & 7.80 & 6.681 & 1.119 & 0.800 \\
3JYA & 6.89 & 5.773 & 1.117 & 0.378 \\
2YGE & 5.06 & 6.173 & 1.113 & 1.000 \\
2YKI & 9.46 & 8.356 & 1.104 & 0.272 \\
4IVC & 10.00 & 8.899 & 1.101 & 0.789 \\
2CBV & 5.48 & 6.580 & 1.100 & 0.333 \\
1YC1 & 6.17 & 7.270 & 1.100 & 0.339 \\
1O3F & 7.96 & 6.862 & 1.098 & 1.000 \\
5C28 & 5.66 & 4.572 & 1.088 & 0.731 \\
4E5W & 7.66 & 8.747 & 1.087 & 0.514 \\
3GR2 & 2.52 & 3.606 & 1.086 & 0.413 \\
4OWM & 2.96 & 4.042 & 1.082 & 0.621 \\
4F2W & 11.30 & 10.223 & 1.077 & 1.000 \\
2XB8 & 7.59 & 6.519 & 1.071 & 1.000 \\
4K18 & 8.96 & 7.891 & 1.069 & 0.537 \\
3COZ & 5.57 & 6.628 & 1.058 & 0.794 \\
3FCQ & 2.77 & 3.819 & 1.049 & 0.559 \\
3G2N & 4.09 & 5.095 & 1.005 & 0.564 \\
3N7A & 3.70 & 4.677 & 0.977 & 1.000 \\
3G2Z & 2.36 & 3.312 & 0.952 & 0.212 \\
3F3D & 7.16 & 6.214 & 0.946 & 1.000 \\
1NC1 & 6.12 & 7.042 & 0.922 & 1.000 \\
3U8K & 8.66 & 7.741 & 0.919 & 0.565 \\
3GE7 & 8.70 & 7.790 & 0.910 & 0.541 \\
3WTJ & 6.53 & 7.440 & 0.910 & 1.000 \\
3COY & 6.02 & 6.925 & 0.905 & 0.836 \\
2HB1 & 3.80 & 4.694 & 0.894 & 0.000 \\
1UTO & 2.27 & 3.164 & 0.894 & 1.000 \\
3TSK & 7.17 & 8.048 & 0.878 & 0.946 \\
1OWH & 7.40 & 6.524 & 0.876 & 0.683 \\
3GNW & 9.10 & 8.234 & 0.866 & 0.635 \\
3ZDG & 7.10 & 6.242 & 0.858 & 0.326 \\
2XII & 7.20 & 8.055 & 0.855 & 0.517 \\
1PXN & 7.15 & 8.002 & 0.852 & 0.558 \\
3ARY & 6.00 & 5.151 & 0.849 & 0.182 \\
4DDK & 2.29 & 3.134 & 0.844 & 0.528 \\
3NW9 & 9.00 & 8.159 & 0.841 & 0.844 \\
3DXG & 2.40 & 3.240 & 0.840 & 1.000 \\
4JFS & 5.27 & 6.099 & 0.829 & 0.297 \\
3UEX & 6.92 & 6.106 & 0.814 & 1.000 \\
4GR0 & 9.55 & 8.738 & 0.812 & 0.829 \\
1P1N & 6.80 & 6.011 & 0.789 & 0.564 \\
3R88 & 4.82 & 4.032 & 0.788 & 0.441 \\
2ZY1 & 7.40 & 6.634 & 0.766 & 0.377 \\
3KR8 & 8.10 & 7.342 & 0.758 & 0.500 \\
4J28 & 5.70 & 6.454 & 0.754 & 0.429 \\
3JVR & 5.72 & 4.983 & 0.737 & 0.321 \\
1LPG & 7.09 & 7.822 & 0.732 & 0.500 \\
4W9H & 6.73 & 6.011 & 0.719 & 0.833 \\
4ABG & 3.57 & 4.289 & 0.719 & 0.676 \\
3AG9 & 8.05 & 7.332 & 0.718 & 0.901 \\
3EBP & 5.91 & 5.196 & 0.714 & 0.297 \\
3PRS & 7.82 & 8.527 & 0.707 & 1.000 \\
1EBY & 9.70 & 8.994 & 0.706 & 0.880 \\
2ZDA & 8.40 & 7.702 & 0.698 & 0.776 \\
4AGP & 4.69 & 5.387 & 0.697 & 0.623 \\
2VVN & 7.30 & 6.607 & 0.693 & 0.667 \\
4MME & 6.50 & 7.186 & 0.686 & 0.218 \\
2ZCR & 6.87 & 7.555 & 0.685 & 0.769 \\
4IVD & 9.52 & 8.847 & 0.673 & 0.750 \\
3PXF & 4.43 & 5.101 & 0.671 & 1.000 \\
1NC3 & 5.00 & 5.663 & 0.663 & 1.000 \\
3JVS & 6.54 & 5.879 & 0.661 & 0.375 \\
4CRA & 7.22 & 7.879 & 0.659 & 0.685 \\
4LZS & 4.80 & 5.458 & 0.658 & 0.473 \\
3OE5 & 6.88 & 7.538 & 0.658 & 0.594 \\
3WZ8 & 5.82 & 6.472 & 0.652 & 0.661 \\
2WNC & 6.32 & 6.963 & 0.643 & 0.444 \\
3O9I & 11.82 & 11.184 & 0.636 & 0.895 \\
3B68 & 8.40 & 9.033 & 0.633 & 0.719 \\
1BZC & 4.92 & 5.549 & 0.629 & 0.569 \\
3GY4 & 5.10 & 4.474 & 0.626 & 0.571 \\
1H22 & 9.10 & 8.474 & 0.626 & 0.220 \\
3QQS & 5.82 & 5.198 & 0.622 & 1.000 \\
1VSO & 4.72 & 5.341 & 0.621 & 0.345 \\
3EHY & 5.85 & 6.469 & 0.619 & 0.558 \\
3GBB & 6.90 & 7.504 & 0.604 & 0.652 \\
3B5R & 8.77 & 8.167 & 0.603 & 0.698 \\
2J78 & 6.42 & 5.817 & 0.603 & 1.000 \\
4GFM & 7.22 & 6.619 & 0.601 & 1.000 \\
4KZQ & 6.10 & 6.694 & 0.594 & 0.564 \\
4DLD & 5.82 & 6.414 & 0.594 & 1.000 \\
4LLX & 2.89 & 3.473 & 0.583 & 0.583 \\
3K5V & 6.30 & 6.883 & 0.583 & 0.322 \\
3SYR & 5.10 & 4.526 & 0.574 & 0.974 \\
1G2K & 7.96 & 8.525 & 0.565 & 0.700 \\
4J21 & 7.41 & 6.847 & 0.563 & 0.543 \\
4E6Q & 8.36 & 7.798 & 0.562 & 0.471 \\
2FXS & 6.06 & 6.595 & 0.535 & 0.295 \\
3OE4 & 7.47 & 6.953 & 0.517 & 0.562 \\
4CRC & 8.72 & 8.205 & 0.515 & 0.591 \\
4F09 & 6.70 & 7.213 & 0.513 & 1.000 \\
3E93 & 8.85 & 8.339 & 0.511 & 0.461 \\
2V7A & 8.30 & 7.791 & 0.509 & 0.340 \\
4JXS & 4.74 & 5.242 & 0.502 & 0.436 \\
2J7H & 7.19 & 6.688 & 0.502 & 1.000 \\
3B27 & 5.16 & 5.660 & 0.500 & 0.340 \\
3ZT2 & 2.84 & 3.335 & 0.495 & 0.625 \\
4BKT & 3.62 & 4.113 & 0.493 & 0.375 \\
3FV1 & 9.30 & 8.819 & 0.481 & 1.000 \\
4DE3 & 5.52 & 5.040 & 0.480 & 0.745 \\
3D4Z & 4.89 & 5.359 & 0.469 & 1.000 \\
4K77 & 6.63 & 7.080 & 0.450 & 0.322 \\
3ARP & 7.15 & 6.706 & 0.444 & 0.246 \\
3ARQ & 6.40 & 5.970 & 0.430 & 0.829 \\
1C5Z & 4.01 & 4.439 & 0.429 & 1.000 \\
2WTV & 8.74 & 8.316 & 0.424 & 0.361 \\
2FVD & 8.52 & 8.121 & 0.399 & 0.277 \\
2QBR & 6.33 & 6.728 & 0.398 & 0.661 \\
3NQ9 & 4.03 & 4.418 & 0.388 & 1.000 \\
2R9W & 5.10 & 4.713 & 0.387 & 0.736 \\
1R5Y & 6.46 & 6.845 & 0.385 & 0.625 \\
2WCA & 5.60 & 5.985 & 0.385 & 1.000 \\
2YFE & 6.63 & 7.012 & 0.382 & 0.833 \\
4DE1 & 5.96 & 5.580 & 0.380 & 0.459 \\
3BGZ & 6.26 & 6.635 & 0.375 & 0.412 \\
3FV2 & 8.11 & 7.744 & 0.366 & 0.729 \\
3ZSX & 3.28 & 3.643 & 0.363 & 0.637 \\
4DJV & 6.72 & 7.065 & 0.345 & 0.673 \\
3UP2 & 7.40 & 7.060 & 0.340 & 0.722 \\
3RLR & 7.52 & 7.854 & 0.334 & 0.660 \\
3DD0 & 9.00 & 8.666 & 0.334 & 1.000 \\
2QBQ & 7.44 & 7.109 & 0.331 & 0.746 \\
1W4O & 5.22 & 4.889 & 0.331 & 1.000 \\
3ARV & 5.64 & 5.970 & 0.330 & 0.660 \\
3UEU & 5.24 & 4.928 & 0.312 & 1.000 \\
4HGE & 7.92 & 8.228 & 0.308 & 0.324 \\
1YDR & 5.52 & 5.826 & 0.306 & 0.491 \\
1GPN & 6.48 & 6.774 & 0.294 & 0.200 \\
4CIG & 3.67 & 3.952 & 0.282 & 0.781 \\
3PYY & 6.86 & 6.579 & 0.281 & 0.227 \\
2QE4 & 7.96 & 7.685 & 0.275 & 0.258 \\
2IWX & 6.68 & 6.952 & 0.272 & 0.976 \\
4IVB & 8.72 & 8.451 & 0.269 & 1.000 \\
3E92 & 8.00 & 7.731 & 0.269 & 0.462 \\
2CET & 8.02 & 8.285 & 0.265 & 1.000 \\
4W9C & 4.65 & 4.914 & 0.264 & 0.793 \\
1O5B & 5.77 & 5.510 & 0.260 & 1.000 \\
3FUR & 8.00 & 7.742 & 0.258 & 0.333 \\
4W9I & 5.96 & 5.705 & 0.255 & 0.746 \\
4OGJ & 6.79 & 6.538 & 0.252 & 0.315 \\
3N76 & 6.85 & 6.600 & 0.250 & 0.440 \\
4DDH & 3.30 & 3.550 & 0.250 & 0.424 \\
2XBV & 8.43 & 8.181 & 0.249 & 0.618 \\
3PWW & 7.32 & 7.562 & 0.242 & 1.000 \\
2QNQ & 6.11 & 6.352 & 0.242 & 0.683 \\
3ZSO & 5.12 & 4.889 & 0.231 & 0.824 \\
2WER & 7.05 & 6.821 & 0.229 & 1.000 \\
4X6P & 8.30 & 8.523 & 0.223 & 0.592 \\
4DLI & 5.62 & 5.843 & 0.223 & 0.364 \\
2WEG & 6.50 & 6.713 & 0.213 & 0.538 \\
1O0H & 5.92 & 5.721 & 0.199 & 1.000 \\
1Z9G & 5.64 & 5.834 & 0.194 & 0.524 \\
1OYT & 7.24 & 7.433 & 0.193 & 0.623 \\
5TMN & 8.04 & 7.848 & 0.192 & 0.513 \\
3P5O & 7.26 & 7.449 & 0.189 & 1.000 \\
4EA2 & 6.44 & 6.619 & 0.179 & 0.288 \\
4M0Y & 6.46 & 6.286 & 0.174 & 0.692 \\
3EJR & 8.57 & 8.406 & 0.164 & 0.761 \\
4GKM & 5.17 & 5.006 & 0.164 & 0.700 \\
4WIV & 6.26 & 6.422 & 0.162 & 0.269 \\
3F3C & 6.02 & 5.860 & 0.160 & 0.724 \\
4F9W & 6.94 & 7.098 & 0.158 & 0.667 \\
3UEV & 5.89 & 5.736 & 0.154 & 1.000 \\
1YDT & 7.32 & 7.168 & 0.152 & 0.607 \\
3UEW & 6.31 & 6.163 & 0.147 & 1.000 \\
2C3I & 7.60 & 7.456 & 0.144 & 0.280 \\
3D6Q & 3.76 & 3.900 & 0.140 & 1.000 \\
1H23 & 8.35 & 8.483 & 0.133 & 0.220 \\
3QGY & 7.80 & 7.928 & 0.128 & 0.253 \\
2WVT & 6.12 & 6.000 & 0.120 & 0.478 \\
1Q8U & 5.96 & 6.076 & 0.116 & 0.318 \\
1K1I & 6.58 & 6.692 & 0.112 & 0.671 \\
3DX2 & 6.82 & 6.711 & 0.109 & 0.429 \\
3CJ4 & 6.51 & 6.402 & 0.108 & 0.719 \\
3GC5 & 7.26 & 7.331 & 0.071 & 0.403 \\
2W4X & 4.85 & 4.783 & 0.067 & 0.615 \\
3E5A & 8.23 & 8.295 & 0.065 & 0.260 \\
3RYJ & 7.80 & 7.862 & 0.062 & 0.778 \\
4CR9 & 4.10 & 4.149 & 0.049 & 0.387 \\
2XYS & 7.42 & 7.374 & 0.046 & 0.184 \\
3TWP & 3.92 & 3.961 & 0.041 & 1.000 \\
1QF1 & 7.32 & 7.280 & 0.040 & 0.766 \\
1NVQ & 8.25 & 8.215 & 0.035 & 0.815 \\
4J3L & 7.80 & 7.769 & 0.031 & 0.688 \\
4EO8 & 8.15 & 8.179 & 0.029 & 0.719 \\
3U5J & 5.61 & 5.584 & 0.026 & 0.608 \\
2VKM & 8.74 & 8.715 & 0.025 & 0.512 \\
3N86 & 5.64 & 5.659 & 0.019 & 0.449 \\
3NX7 & 8.10 & 8.083 & 0.017 & 0.727 \\
4AGQ & 5.01 & 4.999 & 0.011 & 0.647 \\
4KZU & 6.50 & 6.509 & 0.009 & 0.409 \\
4CIW & 4.82 & 4.825 & 0.005 & 0.579 \\
4IH7 & 5.24 & 5.237 & 0.003 & 0.262 \\
\end{longtable}

%\bibliographystyle{abbrv}
%\bibliography{reference}

%\end{document}

\end{document}